\documentclass[iop]{emulateapj}
\usepackage{graphicx}
\usepackage{wrapfig}
\usepackage{accents}

\usepackage{rotating}
\usepackage{epstopdf}
\usepackage{appendix}
\begin{document}

\shorttitle{Circumstellar Dust Around Evolved Stars}
\shortauthors{Villaume et al.}

\title{Circumstellar Dust Around AGB Stars and Implications for Infrared Emission from Galaxies}

\author{Alexa Villaume\altaffilmark{1}, 
Charlie Conroy\altaffilmark{2},  and
Benjamin D. Johnson\altaffilmark{2}} 
\altaffiltext{1}{Department of Astronomy and Astrophysics, University of California, Santa Cruz, CA 95064, USA, avillaum@ucsc.edu} 
\altaffiltext{2}{Harvard-Smithsonian Center for Astrophysics, Cambridge, MA 02138, USA}

\begin{abstract}

  Stellar population synthesis (SPS) models are used to infer many galactic properties including star formation histories, metallicities, and stellar and dust masses.  However, most SPS models neglect the effect of circumstellar dust shells around evolved stars and it is unclear to what extent they impact the analysis of SEDs.  To overcome this shortcoming we have created a new set of circumstellar dust models, using the radiative transfer code DUSTY \citep{ivezic1999}, for asymptotic giant branch (AGB) stars and incorporated them into the Flexible Stellar Population Synthesis code. The circumstellar dust models provide a good fit to individual AGB stars as well as the IR color-magnitude diagrams of the Large and Small Magellanic Clouds. IR luminosity functions from the Large and Small Magellanic Clouds are not well-fit by the 2008 Padova isochrones when coupled to our circumstellar dust models, and so we adjusted the lifetimes of AGB stars in the models to provide a match to the data.  We show, in agreement with previous work, that circumstellar dust from AGB stars can make a significant contribution to the IR ($\gtrsim4\mu m$) emission from galaxies that contain relatively little diffuse dust, including low-metallicity and/or non-star forming galaxies. Our models provide a good fit to the mid-IR spectra of early-type galaxies. Circumstellar dust around AGB stars appears to have a small effect on the IR SEDs of metal-rich star-forming galaxies (i.e., when A$_{\rm V}$ $\gtrsim$~0.1).  Stellar population models that include circumstellar dust will be needed to accurately interpret data from the James Webb Space Telescope (JWST) and other IR facilities.

\end{abstract}


\section{Introduction}

The physical structure, past history, and current properties of a galaxy all go into shaping its observed spectral energy distribution (SED). As such, SEDs are powerful sources of information for unresolved galaxies and have long been used to discern the underlying physical properties of galaxies beginning with the work of \cite{tinsley1972}, \cite{searle1973}, and \cite{larson1978}. These studies pioneered the method of creating synthetic galactic spectra through the sum of the spectra of the stars hosted by the galaxy that has since become known as stellar population synthesis (SPS). By fitting galaxy SEDs with SPS models, properties of the galaxy such as the star formation rate, total mass in stars, metallicity, dust content, and the star formation history can be estimated. For details on the inner workings of SPS models and their broader impact we refer the reader to the recent reviews by \cite{walcher2011} and \cite{conroy2013}.

While there are exciting possibilities for the information we are able to obtain through SPS model fitting, as Charles Babbage noted when first introducing the mechanical computer, we can only expect the right answers if we provide the right input \cite[or, colloquially, ``Garbage in, Garbage Out'';][]{babbage1864}. The limitations of inputs to SPS models are well-known, far ranging, and much discussed. Such limitations include incomplete isochrone tables and stellar libraries, poorly understood stellar evolution, and uncertainties in the initial mass function (IMF). In this paper, we focus on circumstellar dust around asymptotic giant brach (AGB) stars, and the extent to which it affects the integrated light of stellar populations. 

\begin{figure*}[t]
	\includegraphics[width=1.0\textwidth]{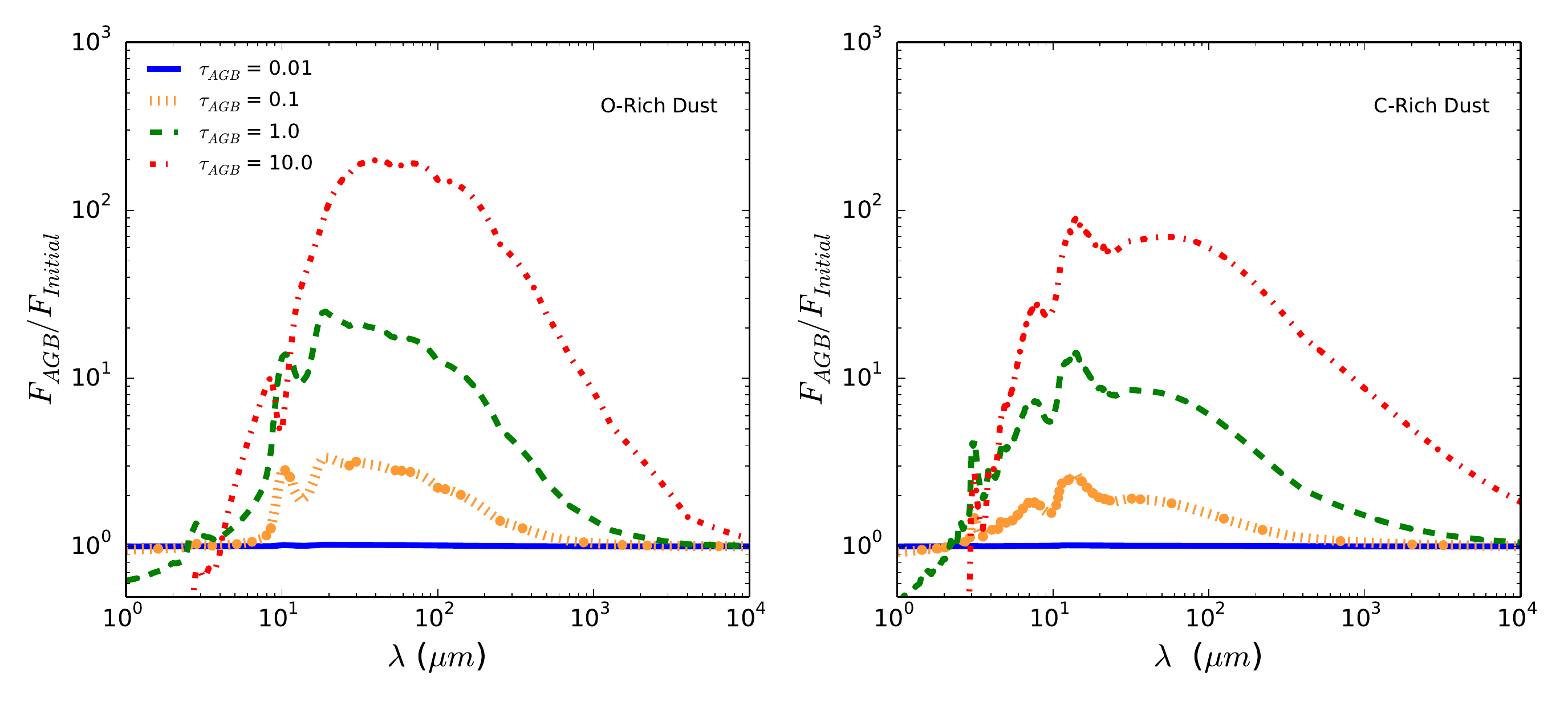}
	\caption{Ratio of the flux including circumstellar dust to the initial input spectra for various values of the dust optical depth at $1\mu m$, $\tau_{\rm AGB}$. The oxygen-rich models (left) are all for $T_{\rm eff}$ = 2000 K while the carbon-rich models (right) are all for $T_{\rm eff}$ = 2400 K. The dominant feature in the oxygen-rich grid is the silicate feature at 10$\mu m$, seen in emission at moderate $\tau_{\rm AGB}$ values and in absorption at high $\tau_{\rm AGB}$ values. The silicon carbide feature at 11$\mu m$ is the dominant feature in the carbon-rich grid which also becomes more prominent for larger values of $\tau_{\rm AGB}$. }
	\label{fig:grid}
\end{figure*}

The AGB phase is the last phase of stellar evolution in intermediate to low mass stars ($\sim 0.1-8M_\odot$) in which significant nuclear burning takes place. During this phase stars eject their envelopes (with mass loss rates up to $10^{-4}M_{\odot}\,{\rm yr}^{-1}$) and evolve toward the white dwarf cooling sequence.   The material around the star is observed to be dust rich (\citealt{bedijn1987}).   This phase is notoriously difficult to model owing to the difficulty in modeling (in 1D) stellar mass loss, convection, and mixing processes in the stellar interior.  AGB stars are very luminous and can contribute many tens of percent to the integrated light of stellar populations \citep{melbourne2013,conroy2013, melbourne2012, kelson2010}.

The helium-shell flashes that occur during the AGB phase trigger the third dredge-up (TDU) process that brings material from the interior of the star to the surface. The surfaces of AGB stars begin as oxygen-rich (C/O $<1$) as a result of the primordial material from which the star forms, but the dredge-up process brings up carbon to the surface and in certain situations results in a carbon-rich envelope (C/O $>1$).  The creation of Carbon stars depends sensitively on the various physical processes shaping the evolution of AGB stars.

Despite these uncertainties, significant progress has been made in modeling AGB stars and their dust envelopes and incorporating those models in SPS models.  On the stellar evolution side, AGB models are becoming increasingly realistic and constrained by observations \citep[see][]{marigo2008, girardi2010, marigo2013,  cassara2013, rosenfield2014}.  There have also been advances in understanding the grain properties of dust around AGB stars from detailed observational studies \citep[e.g.][]{suh1999, suh2000, suh2002, groenewegen2006, groenewegen2009, groenewegen2012, srinivasan2011, sargent2011}. 

The connection between stellar evolution models, which make predictions for the photospheric properties of stars, and the associated dusty circumstellar envelopes is complex and highly uncertain. Nonetheless, several efforts have been made to integrate circumstellar dust models into stellar isochrones and SPS models. \cite{bressan1998} adopted various empirical \citep[or empirically-motivated; see e.g., ][]{vassiliadis1993, habing1994} relations between basic stellar parameters (i.e., mass, luminosity, and radius) and the mass-loss rate, pulsation period, velocity of the ejected material, and the optical depth of the circumstellar envelope. The basic approach of \cite{bressan1998} has been updated and refined by \cite{piovan2003}, \cite{gl2010} and \cite{cassara2013}.

The influence of AGB dust on the SEDs of galaxies is not well understood. \citet{kelson2010} and \citet{chisari2012} suggested that dust from AGB stars may be the dominant contributor to the mid-IR light in star-forming galaxies. However, \citet{melbourne2013} argued that the former studies overstated the effect of AGB stars on galaxies.   In addition, the models of \citet{silva1998} suggest that AGB dust plays a minor role in the SEDs of actively star-forming (and starburst) galaxies.  In old stellar systems, it has been argued that the mid-IR shows evidence for dust around evolved stars \citep[e.g.,][]{athey2002, martini2013}. It is essential to understand these issues because the mid-IR is frequently used as a proxy for the star formation rate (SFR) in galaxies \citep[e.g.,][]{kennicutt2012}. However, the connection between mid-IR flux and SFR may not be quite as simple as commonly assumed, especially for older stellar populations where heating of dust by older stellar populations \citep[e.g.,][]{salim2009, utomo2014}, and emission due to circumstellar dust around evolved stars \citep[e.g.][]{martini2013} may play an important role. At $z\sim2$, the {\it Spitzer} 24$\mu m$ flux probes the rest frame $\sim$8$\mu m$, and so understanding the contributions to the observed SEDs at $\sim$10-30$\mu m$ is critical for deriving reliable SFRs.  In addition, it has been suggested that the mid-IR emission associated with AGB stars could be employed as a stellar population age diagnostic \citep{bressan1998}.  

In this work we create models for the dusty circumstellar envelopes around AGB stars and include those models in the Flexible Stellar Population Synthesis \citep[FSPS][]{conroy2009, conroy2010} models. We employ the publicly available radiative transfer code DUSTY \citep{ivezic1999} to model the dusty envelopes and then an empirically motivated prescription to assign dust shells to AGB stars in isochrones.

The rest of paper is organized as follows: Section 2 details the modeling of circumstellar dust shells and how they are connected to isochrones. In Section 3 we test and calibrate the new dust models using IR data of the Large and Small Magellanic Clouds. In Section 4 we explore the parameter space of the models to see in which regimes the AGB dust has a significant influence on the integrated SEDs of composite stellar populations and compare the models to a variety of extragalactic data. Finally, in Section 5 we discuss the limits and implications of our results.


\section{Including Circumstellar AGB Dust in Stellar Population Synthesis Models}

\subsection{Modeling Dusty Envelopes}

We use the radiative transfer code DUSTY \cite[][]{ivezic1997, ivezic1999} to model dusty circumstellar shells around AGB stars. DUSTY solves the 1D radiative transfer equation for a source embedded in a dusty region assuming spherical symmetry. The input to DUSTY is straightforward --- it requires an input spectrum, the dust properties (chemical composition, grain size distribution, the dust temperature at the inner boundary), a radial density profile, and the optical depth of the envelope at a reference wavelength. We altered the wavelength grid to be finer than the default (1968 wavelength points instead of 105). In this section we describe our choice of inputs and summarize those choices in Table~\ref{table:parameters}.

\begin{deluxetable}{lll}
\tabletypesize{\footnotesize}
\tablecolumns{3}
\tablewidth{0pt}
\tablecaption{Summary of the parameters chosen as input for the DUSTY models \label{table:parameters}}
\tablehead{
\colhead{Parameter} & \colhead{Carbon-Rich}  & \colhead{Oxygen-Rich}}\\
Photosphere Properties &  & \\
\hline
\\
	Photosphere Model  		& Aringer			 & BaSeL	 		 \\
	$T_{\rm eff}$		& $2400-4000$ K 		 & $2000-4000$ K		 \\
	$log~g$				& 0.0 			 & 0.0 				 \\
	Metallicity				& $Z_{\odot}$ & $Z_{\odot}$	 \\
\\
Envelope Properties &  &  \\
\hline
\\
	$R_{\rm in}$ Temperature ($T_c$)	& 1100 K			& 700 K			\\
	Density Profile 		    	& $r^{-2}$	& $r^{-2}$ 	\\
\\
Dust Grain Properties &  & \\
\hline
\\
	$\kappa$ 				& 3200 $\mu m^{2} g^{-1}$		& 3000	$\mu m^{2} g^{-1}$		\\
	$Q_{\rm ext}$		& 0.1 		& 0.1	\\
	Grain Size ($a$)				& 0.1 $\mu m$ 		& 0.1 $\mu m$ 		\\
	Density  ($\rho_d)$		& 2.26 	$g~cm^{-3}$ & 2.5 $g~cm^{-3}$
\end{deluxetable}

We created two separate sets of models, one each for the carbon-rich and oxygen-rich AGB stars. In addition to composition, the models are a function of the optical depth, $\tau_{\rm AGB}$ (herein this refers to optical depth at 1$\mu m$), and the effective temperature of the star, $T_{\rm eff}$. The model DUSTY spectra are implemented in the stellar population synthesis code differentially, in that only the ratio of the output to input spectra are stored and used within the code. 

In Figure~\ref{fig:grid} we show the behavior of the oxygen-rich and carbon-rich grids as a function of optical depth at constant effective temperature for the carbon-rich and oxygen-rich grids. In this figure, we can see the characteristic features of the dust grains used in each set of models. For example, the $11\mu m$ feature caused by SiC in the carbon-rich grid (right) and the $10\mu m$ silicate feature in the oxygen-rich grid (left). The $10\mu m$ silicate feature varies from an emission to an absorption feature as a function of $\tau_{\rm AGB}$.

\begin{figure}[t]
	\includegraphics[height=0.80\textheight]{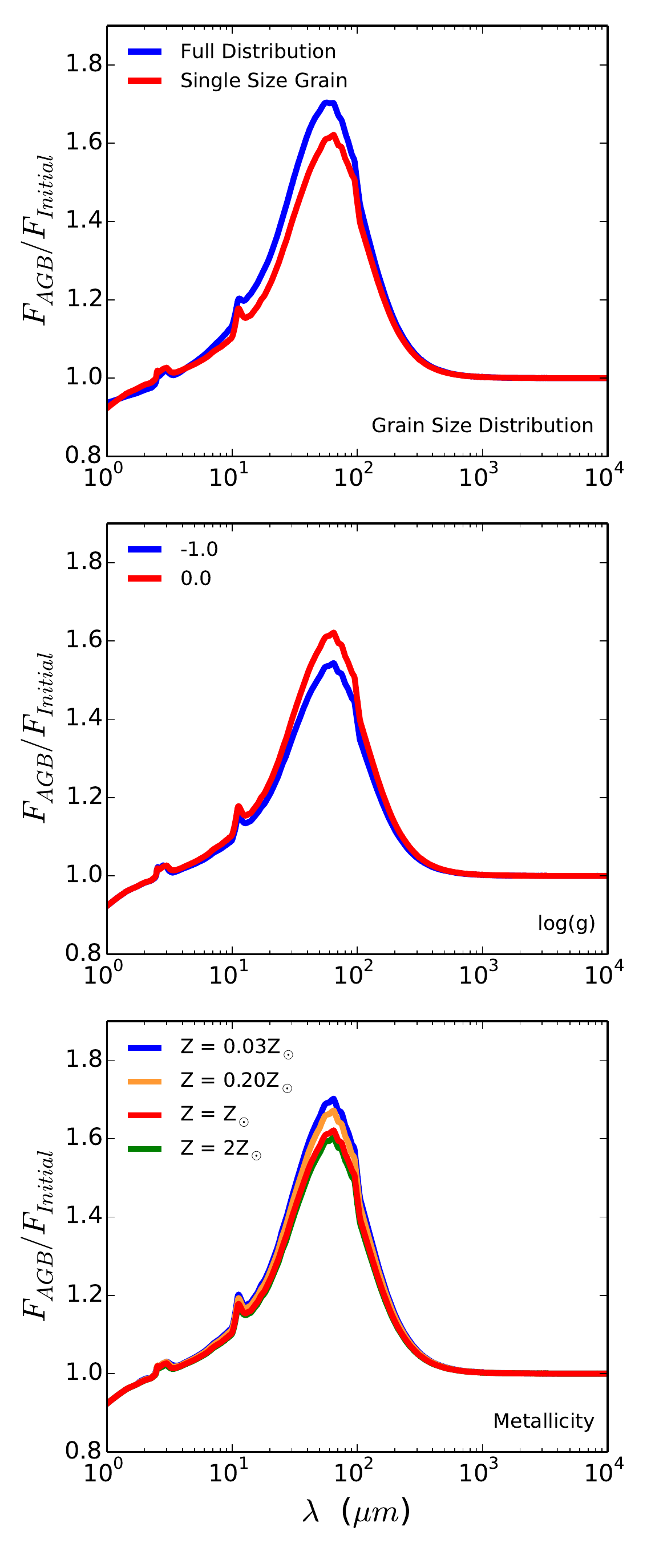}
	\caption{Ratio of the flux including circumstellar dust to the initial input spectra for different grain distributions (top), surface gravity values (middle), and metallicity (bottom) to demonstrate how sensitive the AGB dust models are to these values. The red line in each panel represents our fiducial input spectrum used in the grid with a chosen effective temperature of $3000$ K, all models are for $\tau_{\mathrm AGB} = 0.1$. The differential spectra are generally insensitive to varying the parameters of the input spectrum.}
	\label{fig:input_params}
\end{figure}

The differential implementation of the models allows us to keep the stellar parameters beyond effective temperature, such as stellar metallicity and surface gravity, constant. In Figure \ref{fig:input_params} we show the sensitivity of the differential spectra to surface gravity (middle panel) and metallicity (bottom panel). As can been seen, variations of these parameters does not have a significant impact on the differential spectrum.  As such, surface gravity is kept constant at log(g)$=0.0$ and the metallicity at $Z=Z_{\odot}$ for both the carbon-rich and oxygen-rich differential spectra.

For the oxygen-rich input stellar spectra we use the BaSeL (\citealt{lejeune1997}) spectral library with effective temperatures ranging from 2000 K to 4000 K. For the carbon-rich models we use the Aringer (\citealt{aringer2009}) spectral library with effective temperatures ranging from 2400 K to 4000 K. Stars hotter than $\sim4000$K are not expected to contain significant circumstellar dust shells.

A key characteristic of the dust is quantified through the value of the extinction coefficient, $\kappa$,

\begin{equation}
	\kappa = \frac{n_\mathrm{d} \pi a^2Q_\mathrm{ext}}{\rho_\mathrm{d}},
	\label{eq:kappa}
\end{equation}

\noindent where $Q_{\rm ext}$ is the extinction efficiency factor, $a$ is the grain size, $\rho_d$ is the internal grain density, and $n_d$ is the spatial number density of a particular grain species in a dust mixture.  The final $\kappa$ for each grain mixture is the sum of the $\kappa$ values for each component in that mixture.  We fix $Q_{\rm ext}=0.1$ throughout, motivated by \cite{suh1999}, and the grain densities are adopted from \cite{bressan1998} (See Table \ref{table:parameters}).

\begin{deluxetable}{lll}
\tablecaption{Dust condensation temperature ($T_c$) }
\tablehead{
\colhead{} & \colhead{Carbon-Rich}  & \colhead{Oxygen-Rich}
}\\
\startdata
This work & 1100 K & 700 K \\
Cassara et al. (2013) & 1500 K & 1000 K \\ 
Groenewegen et al. (2009a) & 900-1200 K & 900-1200 K \\
Groenewegen et al. (2009b)\tablenotemark{a} & 1000-1200 K & 800-1000 K\\
Marigo et al.  (2008) & 800-1500 K & 800-1500 K \\
Groenewegen (2006) & 1000-1200 K & 1000-1500 K\\
Piovan et al. (2003) & 1000 K & 1000 K \\
Lorenz-Martins \& Pompeia (2001)\tablenotemark{a}& - &  417 - 1011 K\\ 
Suh (2000) & 1000 K &  - \\
Suh (1999) & - & 700/1000 K \tablenotemark{b} \\
Bressan et al.  (1998) & 1500 K & 1000 K \\
David \& Papoular (1990) & - & 500 K - 800 K \\
Rowen-Robinson \& Harris (1982) & - & 500/1000 K\\ 
\enddata
\tablenotetext{a}{Values were fitted as part of a model}\tablenotetext{b}{\citet{suh1999} mentions that both temperatures have been used in the literature but chooses 1000 K for their own models}
\label{table:rin}
\end{deluxetable}

Throughout this work we adopt a grain size distribution that is a delta function at 0.1$\mu m$. This is a common, though rather simplistic assumption \citep[e.g.,][]{suh1999, suh2002, piovan2003}. We tested the impact of adopting different grain size distributions in the DUSTY models and found that we can obtain similar emergent SEDs for different grain size distributions by varying $\tau_{\rm AGB}$ (see top panel of Figure \ref{fig:input_params}).

In our model the parameters $\kappa$, the dust-to-gas ratio, and $R_{\rm in}$ depend on whether the star is oxygen-rich or carbon-rich.  To compute $\kappa$ we need to assume a certain dust composition. It is not currently possible to compute $\kappa$ from first principles owing to various uncertainties such as the detailed properties and evolution of circumstellar dust shells and dust grain formation and destruction processes, although efforts along these lines are currently ongoing (e.g., \citealt{jones2012}, \citealt{ventura2014}, \citealt{nanni2013}, \citealt{schneider2014}, \citealt{dell'agli2014}).

 As \cite{cassara2013} discuss in detail, we expect the dust composition to change as a function of $\tau_{\rm AGB}$. In principle, this would require us to compute $\kappa$ and $\tau_{\rm AGB}$ iteratively (\citealt{piovan2003}). However, AGB stars can have diverse features from one another and it is unlikely that any one set of uniform models will fit all AGB stars equally. Since we have no a priori model for how the grain properties should vary with stellar properties, we choose to adopt a relatively simple, observationally motivated scheme. 

\begin{figure*}[t]
	\includegraphics[width=1.0\textwidth]{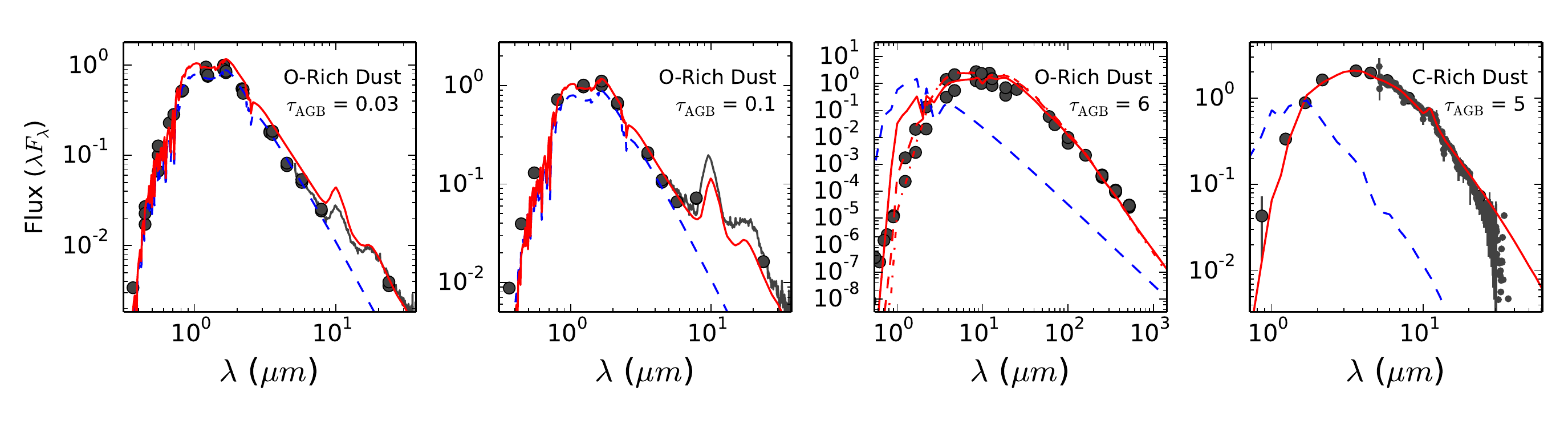}
	\caption{By-eye fits to four AGB stars using radiative transfer models. The blue dashed line is the input stellar photosphere and the red line shows the effect of including a circumstellar dust shell. The normalization in each panel is arbitrary. Far left: M star HV5715 with $T_{\rm eff}$ = 3500K and a grain composition off 100\% warm silicates. Middle left: M star SSTISAGE1C J052206.92 with $T_{\rm eff}$ = 3500K and  grain composition of 100\% warm silicates. Middle right: M star CW Leo with $T_{\rm eff}=2000$K; this star is very dust enshrouded and so we adopt a grain composition of 100\% cold silicates. Different lines are plotted showing different values of the dust temperature at the inner boundary, $T_{c}$. We show $T_{c}$ = 700K, 1000K, 1300K (see text for details). There is a general agreement of the models for  different $T_{c}$ values especially at the mid-IR wavelengths. Far right: Carbon star LPV 28579 with $T_{\rm eff}$ = 3200K.}
	\label{fig:model}
\end{figure*}


\cite{sargent2010} and \cite{srinivasan2010} fit models to multi-wavelength broadband photometry of AGB stars with grain composition as a free parameter of the model. For oxygen-rich stars, \cite{sargent2010} found that a grain composition of 100\% oxygen-deficient silicates was sufficient to fit the data.  However, as an oxygen-rich shell becomes more optically thick we expect the silicates to become colder (\citealt{suh1999} and \citealt{suh2002}).  Therefore, in our grid we adopt a grain type that varies with $\tau_{\rm AGB}$.  In our grid, for oxygen-rich stars we use warm silicates for the dust composition up until $\tau_{\rm AGB}$ = 3 when we start to include cold silicates in the mixture. The grain densities, sizes, and $Q_\mathrm{ext}$ parameters are all kept constant between the warm and cold silicates. For this reason, the extinction coefficient (Equation 1) remains the same as the grid transitions from cold to warm silicates.

For carbon-rich stars, \cite{srinivasan2010} found that a mixture of amorphous carbon (amC) and 10\% silicon carbide (SiC) was sufficient. These dust compositions were used for their entire grid of AGB dust models, GRAMS (\citealt{sargent2011} and \citealt{srinivasan2011}). Furthermore, we found that changing the amC and SiC ratio made only a modest difference in our results so we decided to keep the grain mixture for carbon-rich stars constant at 90$\%$ amC and 10$\%$ SiC for all values of $\tau_{\rm AGB}$. 

Despite the simplicity of our dust compositions we find that they are sufficient for modeling individual AGB stars. We demonstrate this in Figure~\ref{fig:model} where we show how the DUSTY models compare to the SEDs of well-studied AGB stars HV 5715, SSTISAGE1C J052206.92, LPV 28579, and CW Leo. For the first three stars the data is as presented in \citet{sargent2010} and \citet{srinivasan2010} with  {\it UBVI} \citep[][Magellanic Clouds Photometric Survey]{zaritsky1997}, {\it JHK} \citep[][2 Micron All Sky Survey]{Skrutskie2006}, {\it Spitzer} IRAC and MIPS bandpasses \citep[][SAGE]{meixner2006}, and {\it Spitzer} IRS spectroscopy \citep[][SAGE-Spec]{kemper2010}. CW Leo is an extensively observed object. Here we fit optical to IR photometry \citep[as presented in][]{groenewegen2012} -- {\it gri} \citep[][Sloan Digital Sky Survey]{ahn2012}, {\it VRI} \citep[][]{lebertre1987}, {\it JHKLM} \citep[][]{lebertre1992}, IRAS \citep[][]{beichman1988}, AKARI  \citep[][AKARI IRC Survey of the Large Magellanic Cloud]{ita2008}, and SPIRE and PACS bandpasses \citep[][Mass-Loss of Evolved StarS]{groenewegen2011}.

HV 5715,  SSTISAGE1C J052206.92, LPV 28579, and CW Leo are all oxygen-rich AGB stars and LPV 28579 is carbon-rich. \cite{sargent2011} presented detailed models of HV 5715 and SSTISAGE1C J052206.92, \cite{srinivasan2011} modeled LPV 28579, and \citet{groenewegen2012} modeled CW Leo. The models used in Figure~\ref{fig:model} were made using the default parameters of our model grid and only allowing $T_{\rm eff}$ and $\tau_{\rm AGB}$ to vary. The shape of the observed SEDs and the observed features are fit well by the models. This comparison provides a test of our model grid both as a function of $T_{\rm eff}$, $\tau_{\rm AGB}$, and whether the star is oxygen or carbon-rich.  We emphasize however that these fits are not unique and there are many degeneracies between e.g., the shell density profile, $\tau_{\rm AGB}$, grain size distribution, etc.

The temperature of the dust at the inner radius, $R_{\rm in}$, of the shell, $T_{c}$, is an important source of uncertainty in the modeling. Table~\ref{table:rin} summarizes the different values for $T_{c}$ adopted in the literature and demonstrates the range of values considered by various authors.  Our choice for $T_{c}$ was based on fitting the data from \cite{martini2013} (as presented in Section 4). 

We compared the dust models of individual stars for the different $T_{c}$ values to the stars shown in Figure \ref{fig:model} to check that on an individual basis the values gave reasonable agreement to AGB stars. In Figure \ref{fig:model} we compare the different models for the range of $T_{c}$ values to CW Leo. Since CW Leo is the most dust enshrouded AGB star of our sample any effects due to changes to the input parameters will be amplified compared to the other stars in the sample. These comparisons lead us to adopt $T_{c}=700$ K for the oxygen-rich grid.  We emphasize that $T_c$ could in principle vary with stellar type (i.e., with $T_{\rm eff}$), metallicity, etc., and our choice of a constant value for $T_c$ may introduce systematic uncertainties in the final model results. 

The grid and the code used to generate the grid are publicly available\footnote{https://github.com/AlexaVillaume/AGBGrid \\
commit: 7308b5c424268c16da3e4aeee9aef0b5f0bfcaf6 }.

\subsection{Connecting Dusty Envelopes to Stellar Isochrones}

\begin{figure*}[ht]
  \center
	\includegraphics[width=0.9\textwidth]{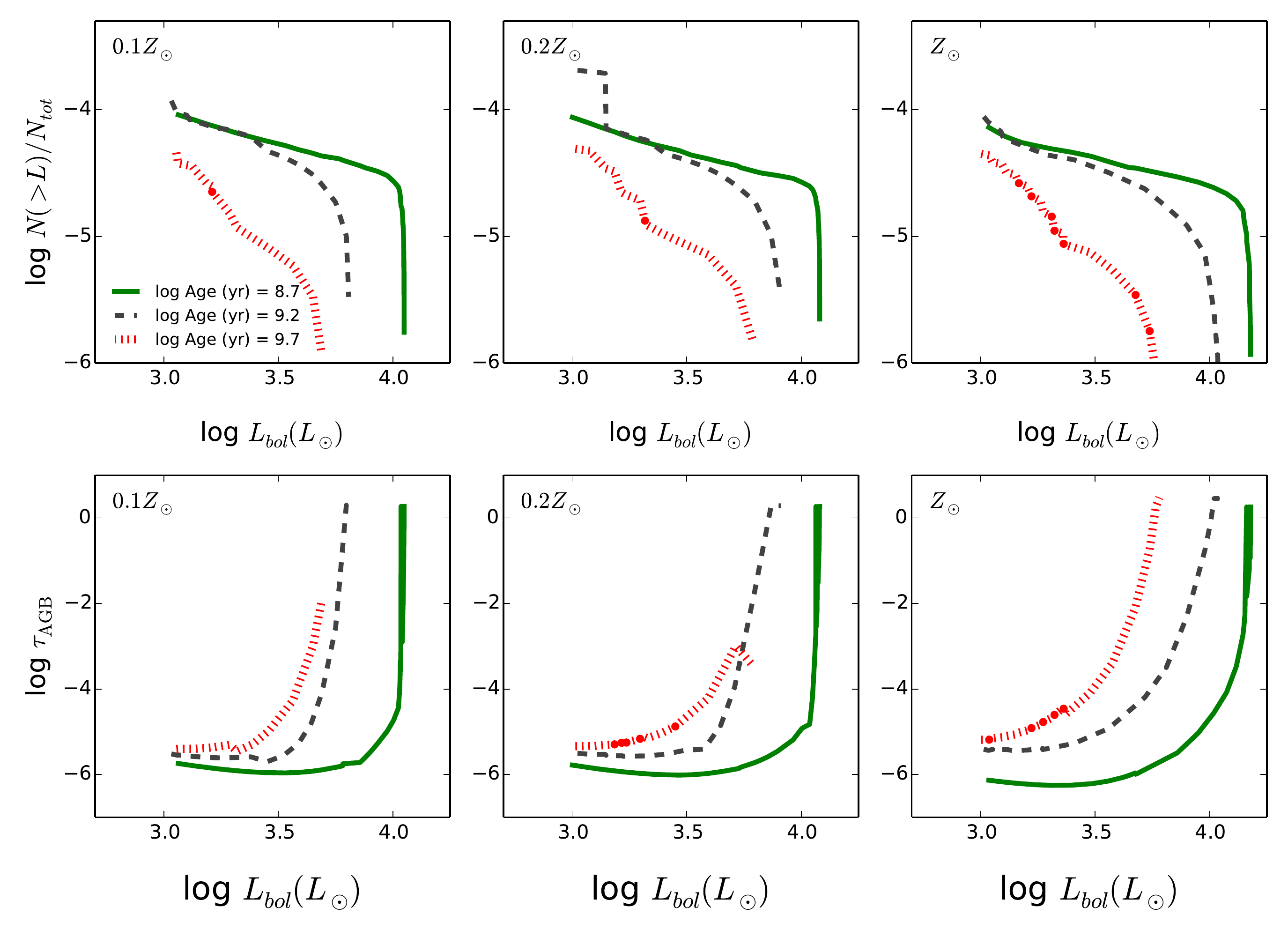}
	\caption{Top Panels: Cumulative luminosity functions for AGB stars for different ages and metallicities. Bottom Panels: Value of $\tau_{\rm AGB}$ as a function of L$_{\rm bol}$, age, and metallicity.  Only stars with $L_{\rm bol} \ge 10^3 L_{\odot}$ are shown. Stars that have the highest $\tau_{\rm AGB}$ values, i.e. the most dust-enshrouded, have the greatest bolometric luminosity, but are rare.}
	\label{fig:lum_func}
\end{figure*}

With grids of circumstellar AGB dust emission as a function of optical depth, $T_{\rm eff}$, and C/O taken from the dustless isochrones, our goal now is to connect the grids to stellar isochrones by computing the value of $\tau_{\rm AGB}$ at each isochrone point.

The optical depth, $\tau_{\rm AGB}$, is the key quantity that connects the stellar parameters provided by the isochrones to the models of the dust shells. Computing $\tau_{\rm AGB}$ from stellar parameters has been discussed extensively in the literature beginning with the work of \cite{vassiliadis1993} and \cite{habing1994}. Subsequent work by  \cite{bressan1998}, \cite{piovan2003}, and \cite{cassara2013} refined this technique to incorporate dusty envelopes into their SPS codes. In this work, we largely follow these previous efforts in coupling circumstellar dust models to stellar isochrones. We will be brief in our description of computing $\tau_{\rm AGB}$ as the former three papers provide extensive details of the derivations of the following equations.

We start with an initial equation for $\tau_{\rm AGB}$ that is derived assuming spherical symmetry and by integrating over the thickness of the shell while assuming that $R_{\rm out}\gg R_{\rm in}$, where  $R_{\rm out}$ and $R_{\rm in}$ are the outer and inner radii of the dust envelope:

\begin{equation}
	\tau_{\rm AGB} = \frac{\delta \dot{M} \kappa}{4 \pi v_\mathrm{exp}} \frac{1}{R_{\rm in}}.
	\label{eq:tau}
\end{equation}

In \cite{vassiliadis1993} they used observationally estimated mass-loss rates of Galactic Mira variables and OH/IR stars to empirically determine equations for $\dot{M}$, pulsation period ($P$), and $v_\mathrm{exp}$:

\begin{equation}
	\frac{v_\mathrm{exp}}{\mathrm{km}~s^{-1}} = {-13.5 + 0.056\frac{P}{\mathrm{days}}}, 
	\label{eq:vexp}
\end{equation}

\noindent with an additional condition that $v_\mathrm{exp}$ lie in the range of $3-15$ km s$^{-1}$, and:

\begin{equation}
	\mathrm{log}~\frac{P}{\mathrm{days}} = -2.07 + 1.94~\mathrm{log}~\frac{R}{R_\mathrm{\odot}} - 0.9~\mathrm{log}~\frac{M}{M_\mathrm{\odot}}.
	\label{eq:pulse}
\end{equation}

The equations for $\dot{M}$ are a function of initial mass of the star and whether the star is in a super-wind phase, defined by \cite{vassiliadis1993} as stars with $P>500$ days.  When not in the super-wind phase the mass-loss rate for a star with an initial mass $<2.5M_{\odot}$ is described by a simple relation with pulsation period,

\begin{equation}
	\mathrm{log}~\dot{M} = -11.4 + 0.0123P, 
	\label{eq:m1}
\end{equation}

\noindent where $P$ is in days, and $\dot{M}$ is in $M_{\odot}\,{\rm yr}^{-1}$, and for stars with initial mass $>2.5 M_{\odot}$,

\begin{equation}
	\mathrm{log}~\dot{M} = -11.4 + 0.0125 \left[P - 100\left( \frac{M}{M_\mathrm{\odot}} - 2.5\right) \right] .
	\label{eq:m2}
\end{equation}

\noindent In the super-wind phase the mass-loss rate is described as,  
\begin{equation}
	\dot{M} = \frac{1}{c\,v_{exp}}\frac{L}{L_\mathrm{\odot}}.
	\label{eq:m3}
\end{equation}

\noindent where $c$ and $v_{exp}$ are in units of km s$^{-1}$.

The inner radius, $R_{\rm in}$, is derived by equating the stellar luminosity with the luminosity at the inner radius, remembering that the inner radius is defined as the radius where the temperature is equal to the dust condensation temperature, $T_{c}$,
\begin{equation}
	R_{\rm in} = \left [ \frac{L}{4 \pi T^4_\mathrm{c} } \right ]^{ \frac{1}{2}} .
	\label{eq:rin1}
\end{equation}

\noindent We adopt our $T_{c}$ values from Table~\ref{table:rin} to obtain a final relation for carbon-rich stars,

\begin{equation}
	R_{\rm in} = 1.92 \times 10^{12}\left( \frac{L}{L_\mathrm{\odot}} \right)^{\frac{1}{2}} {\rm cm},
\end{equation}

\noindent and for oxygen-rich stars,

\begin{equation}
	R_{\rm in}  = 4.74 \times 10^{12} \left(  \frac{L}{L_\mathrm{\odot}} \right)^{\frac{1}{2}} {\rm cm}.
\end{equation}

The dust-to-gas ratio, $\delta$, is obtained by inverting the observed correlation between it and $v_\mathrm{exp}$ found in \cite{habing1994}, 

\begin{equation}
	\delta = \frac{\delta_{\rm AGB} \, v_\mathrm{exp}^2}{225} \left(\frac{L}{10^4 L_\mathrm{\odot}}\right)^{-0.06},
	\label{eq:delta}
\end{equation}

\noindent where we adopt $\delta_{\rm AGB}$ = 0.01 for oxygen-rich stars \citep{suh1999} and $\delta_{\rm AGB}$ = 0.0025 for carbon-rich stars \citep{blanco1998}. Note that in the following we assume that $\delta$ does not depend on metallicity. See Section 2.3 for details.

With the above equations we can now explore the behavior of $\tau_{\rm AGB}$ along the isochrones as a function of metallicity and age.  This is shown in Figure~\ref{fig:lum_func} where we show the number of AGB stars as a function of luminosity (top) and their corresponding $\tau_{\rm AGB}$ values (bottom).  In this figure we see that the knee of the luminosity functions coincide with the upturn of the $\tau_{\rm AGB}$ values. This indicates that brighter stars are both rarer and have larger values of $\tau_{\rm AGB}$. Only for these rare stars is there an appreciable level of dust obscuration.
 
The models are now included in the FSPS population synthesis code (\citealt{conroy2010}) as of v2.5.\footnote{\texttt{code.google.com/p/fsps}} FSPS computes $\tau_{\rm AGB}$ as per the equations above for identified AGB stars in the isochrones. With $\tau_{\rm AGB}$, $T_{\rm eff}$, and the composition of the stellar envelope, FSPS interpolates within the grid of DUSTY models and grafts the interpolated model onto the stellar SED.   

We briefly review here the salient characteristics of the FSPS stellar population synthesis model.  FSPS takes as input a set of stellar isochrones (here we use the Padova models detailed in \cite{marigo2008}), stellar spectral libraries (in our case the BaSeL library), allows the user to specify a stellar initial mass function (IMF), and outputs simple stellar populations (SSPs), and, if a star formation history is specified, composite stellar populations.  FSPS includes models for diffuse dust absorption and emission.  The model has been extensively calibrated and tested against observations, see \cite{conroy2010} for details.

When AGB dust is turned on in FSPS, $\tau_{\mathrm AGB}$ is computed at each isochrone point using the method described above. Fundamentally, $\tau_{\mathrm AGB}$  depends on the mass and age of the star.

\subsection{Metallicity Dependence of the Dust-to-Gas Ratio?}

To first order, the seed elements relevant for grain formation in oxygen-rich environments, such as O, Si, and Fe, will vary in proportion to the metallicity of the star. It therefore might seem reasonable for the dust-to-gas ratio to increase linearly with the metallicity of the star \citep[e.g.,][]{habing1994}.  

However, recent models attempting to follow the formation and dust grains self-consistently as a function of AGB mass and metallicity have revealed a complicated relation between the dust-to-gas ratio and metallicity \citep[e.g.][]{ferrarotti2006}.  \cite{schneider2014} detailed the complicated factors that contribute to the dust production of AGB stars by comparing the observed dust production rates of AGB stars in the Large and Small Magellanic Clouds to theoretical models. \cite{schneider2014} noted that efficiency of the Hot Bottom Burning (HBB), the TDU, and the mass of the star all contribute to the efficiency of the dust production for AGB stars. And, as \cite{ventura2014} noted, the efficiency of these mechanisms are influenced by the modeling of convection in the star. Discrepancies between the theoretical dust yields and observed dust yields found in \cite{schneider2014} highlight that we do not have a solid predictive model for dust formation in AGB winds. Moreover, current models suggest that dust production could be a complex, non-monotonic function of metallicity. For these reasons we decided to adopt a simple approach and keep the dust-to-gas ratio fixed for all oxygen-rich stars, independently of metallicity.

For carbon-rich stars the seed elements for the grains (carbon) are made in-situ in the star, so one would expect that the dust-to-gas ratio to have little dependence on the original metallicity of the star.  Hence also for carbon-rich stars we keep the dust-to-gas ratio fixed, independent of metallicity.


\section{Calibrating Stellar Population Models That Include AGB Dust}

\begin{figure}[t!]
	\includegraphics[width=0.5\textwidth]{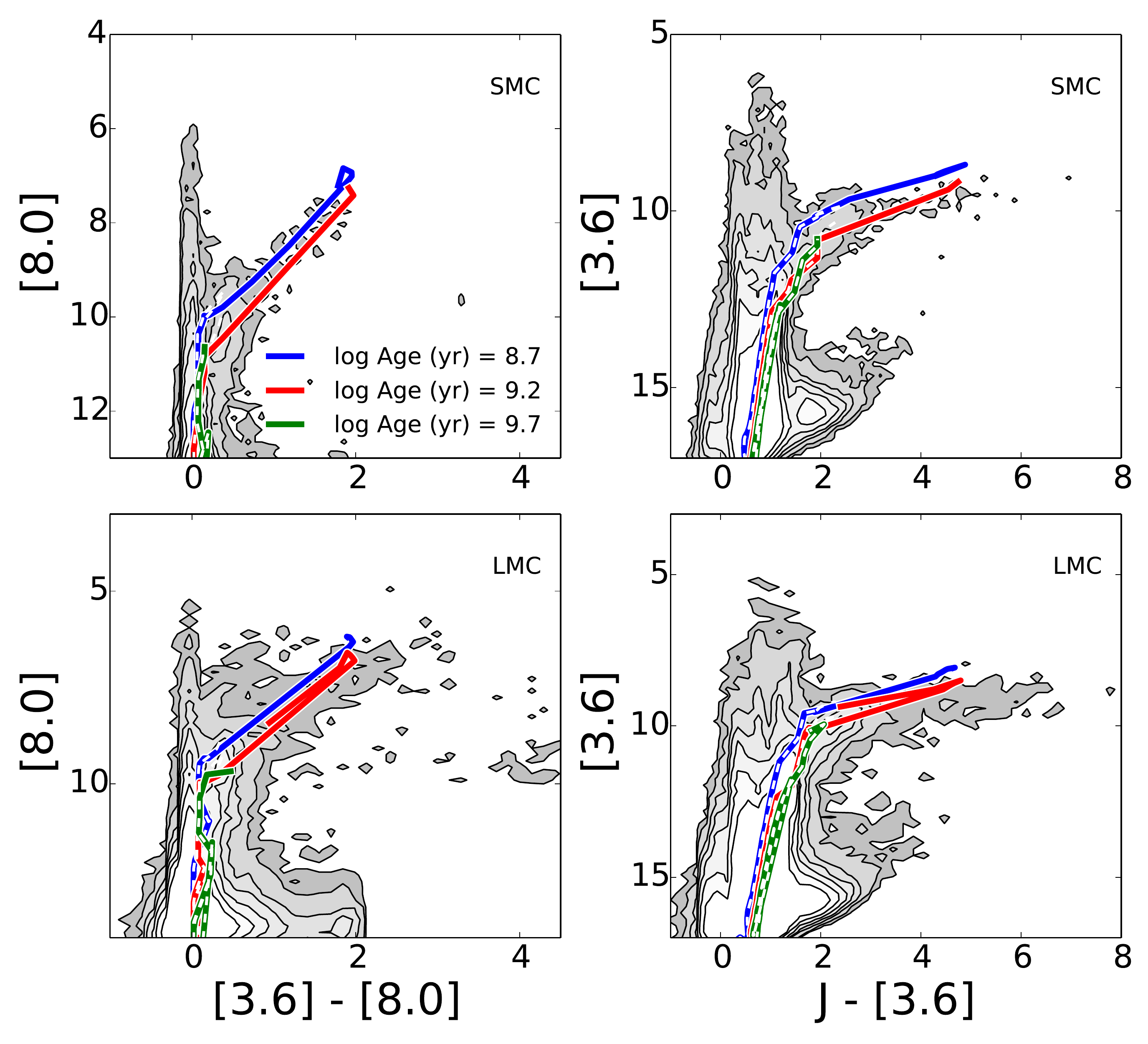}
	\caption{Color magnitude diagrams of stars in the Small Magellanic Cloud (top) and the Large Magellanic Cloud (bottom) overlaid with SSP models of various ages. Foreground objects have been removed with a color-magnitude cut. The solid lines are single-age models that include AGB circumstellar dust  while the dashed lines show the same models without AGB dust. Note that the luminous red spur in the data is only captured by the models that include AGB dust.}
	\label{fig:lmcsmc}
\end{figure}

As an initial test of the new models, we compare them with the photometry of the Large Magellanic Cloud (LMC) and the Small Magellanic Cloud (SMC). The data is from the Survey of the Agents of Galaxy Evolution (SAGE) survey (\citealt{meixner2006} for LMC data and \citealt{gordon2011} for SMC). The SAGE survey is based on Spitzer IRAC and MIPS data, including photometry in [3.6], [4.5], [5.8], and [8.0] bands.  

\begin{figure}[t!]
	\includegraphics[width=0.5\textwidth]{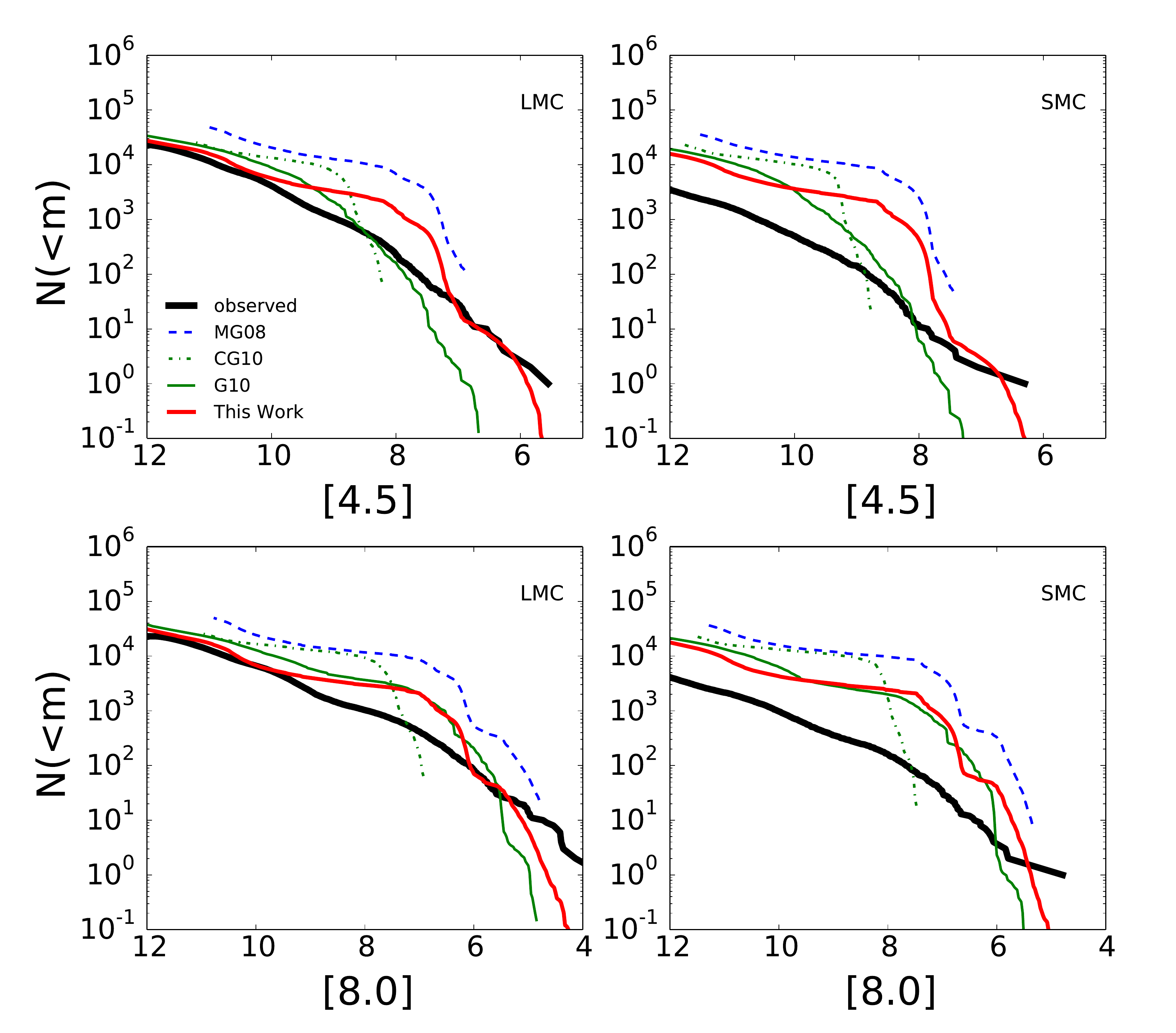}
	\caption{Comparison of the observed and model luminosity functions for the SMC and LMC.  Model LFs were derived using the default Padova 2008 isochrones \citep{marigo2008}, the updated Padova isochrones by \citet{girardi2010} (Case A), isochrones as calibrated by \citet{conroy2010}, and isochrones calibrated for this work.  All model predictions include circumstellar dust for AGB stars and have been convolved with estimated SFHs of the LMC and SMC. }
	\label{fig:lf}
\end{figure}

For the LMC data we follow \cite{cassara2013} and select a region within $\pi$ square degrees of the center of the LMC (RA = 5h23m.5, DEC = -69,45'). For the [3.6] - [8.0] vs [8.0] color-magnitude diagram (CMD) for both galaxies we make a further cut on the dim, red background objects for clarity. In Figure~\ref{fig:lmcsmc} we include isochrones for both the LMC and SMC, with respective metallicities of $Z=0.008$ and $Z=0.004$, from the Padova stellar models.

In Figure~\ref{fig:lmcsmc} there are six models in each panel: three ages, and for each age there are models both with and without AGB dust (solid and dashed lines, respectively). Overall, the models with circumstellar dust cover the observed behavior of both the LMC and SMC, providing support to our method for assigning dusty envelopes to AGB stars in isochrones (see also \citet{cassara2013} who find similar agreement with a different set of AGB dust models).

Comparing isochrones to data in CMD space hides potentially significant discrepancies in the relative numbers of stars in different evolutionary phases. We provide a closer comparison of our models to LMC and SMC luminosity functions (LFs) in Figure~\ref{fig:lf}. The observed cumulative LFs (thick black line in Figure~\ref{fig:lf}) are constructed from the catalogs of \citet{boyer2011}, including only those stars classified as any variety of O, C, or x-AGB. We have inspected images of the brightest objects to insure that they are not strongly affected by confusion or high backgrounds. 

We create model LFs by convolving model isochrones with measured SFHs and metallicity histories from \citet{harris2004} and applying a cut of $T_{\rm eff}<4000$ K to isolate the AGB stars. We checked that the convolution of the SFHs with the stellar mass and integrated luminosities reproduces the observed integrated NIR light of the clouds and the total stellar masses reported by \citet{harris2004} and \citet{harris2009} to within a factor of two. We show four models in Figure~\ref{fig:lf}, three of which differ only in the input isochrones: unaltered Padova isochrones \citep{marigo2008} (MG08, dashed blue line), re-calibrated Padova isochrones from \cite{conroy2010} (CG10, dot-dashed green line), and a new normalization created specifically to provide a better fit to the observed LF (sold red line, described in detail below). The fourth is computed from the updated Padova isochrones as described in \citet{girardi2010} (Case A). We emphasize that this is an entirely different model where we adopt the \citet{girardi2010} bolometric corrections and \citet{marigo2008} dust models. In all cases circumstellar dust around AGB stars has been included.

\begin{figure}[t!]
	\includegraphics[width=0.5\textwidth]{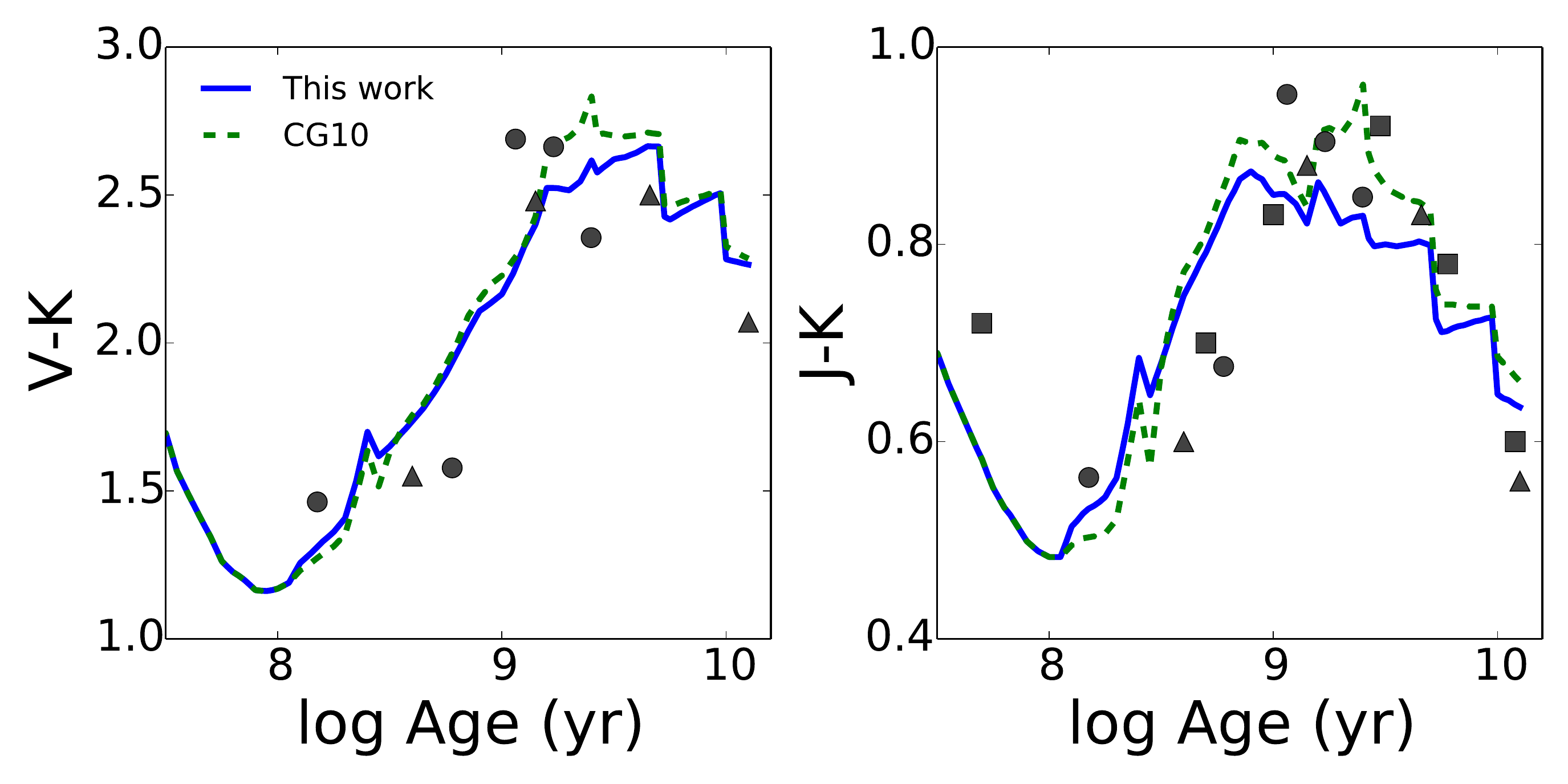}
	\caption{SSP colors as a function of age, comparing models from this work to those described in \citet{conroy2010}.  This figure also includes LMC star cluster data from \citet{noel2013} (circles), \citet{pessev2008} (triangles), and \citet{gonzalez2004} (squares). By construction, the colors predicted by the calibrations described in this work are very similar to the colors from \citet{conroy2010}.  Notice that the very similar integrated light colors between the two models nonetheless produce very different LFs in Figure \ref{fig:lf}.}
	\label{fig:int_light}
\end{figure}

\begin{figure*}[t]
	\includegraphics[width=1.0\textwidth]{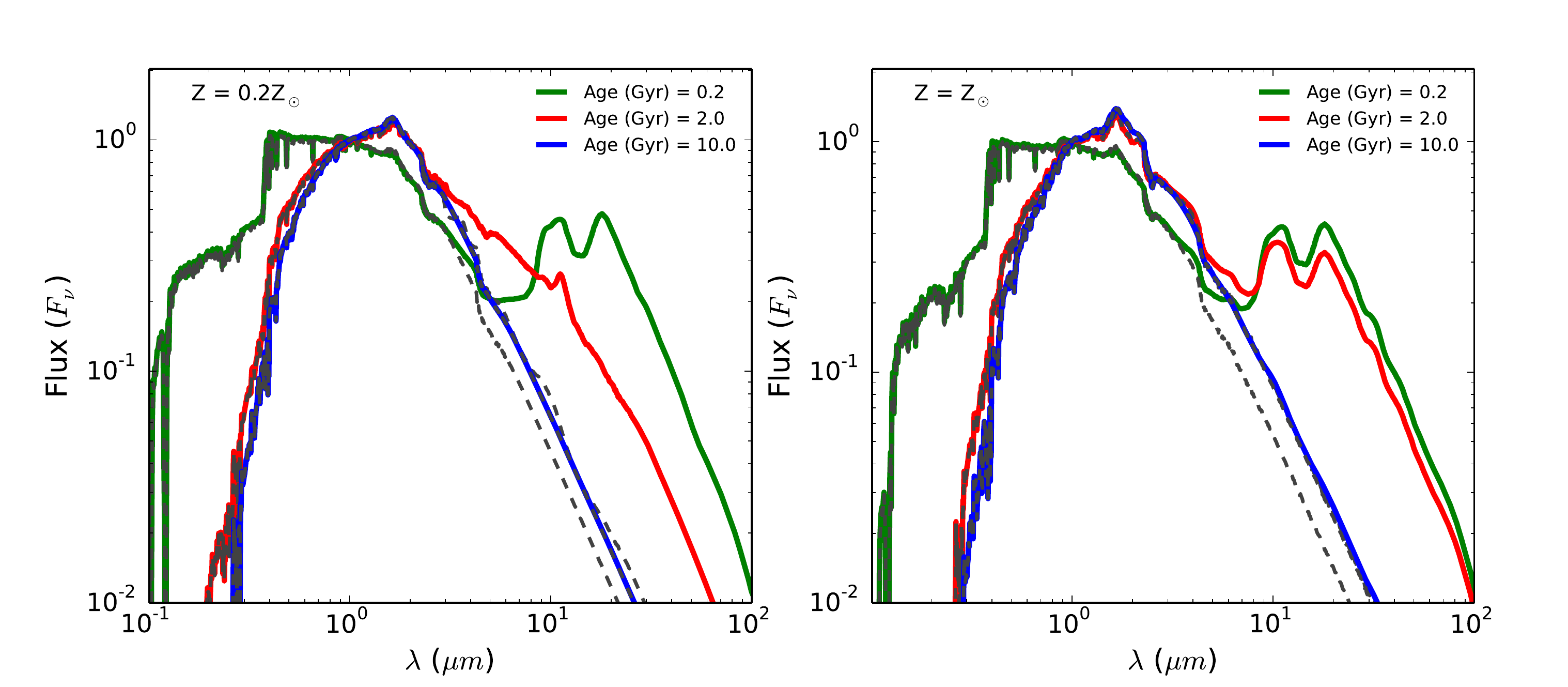}
	\caption{Model spectral energy distributions as a function of age comparing models with (solid lines) and without (dashed lines) circumstellar dust around AGB stars.  Notice that circumstellar dust has a very large effect on the SED for $\lambda\gtrsim4\mu m$.  For the Padova stellar models that we employ, the effect of circumstellar dust around AGB stars is relatively modest at 10 Gyr.}
	\label{fig:sed}
\end{figure*}

Uncertainties in the SFHs are difficult to incorporate correctly due to the substantial covariance between the SFRs in adjacent temporal bins, but the uncertainties on the SFR in any given bin are typically less than a factor of two, and this gives an upper limit on the uncertainties in the predicted CLFs at any magnitude.  The effects of photometric uncertainties on the predicted CLFs are negligible, though uncertainties substantially larger than reported may affect the bright end.

The predicted luminosity functions determined from the \cite{marigo2008} isochrones and the \cite{conroy2010} calibrations do not match the data well. Models based on the \cite{marigo2008} isochrones over-predict the number of AGB stars while models based on the \cite{conroy2010} normalization do not span the full extent of the magnitude range. Due to these shortcoming we decided to introduce updated calibrations of the Padova isochrones for the AGB phase, different from those presented in \cite{conroy2010}.

\citet{conroy2010} chose to decrease the overall contribution of AGB stars to the integrated light by decreasing log $L$ and slightly modifying $T_{\rm eff}$.  This modification resulted in a better fit to the NIR colors of intermediate age star clusters in the LMC.   For the new calibration, we chose to keep log $L$ unchanged and instead adjust the IMF weight to effectively reduce the lifetimes of AGB stars:

\begin{equation}
	\mathrm{weight} = \mathrm{max}(+0.1, 10^{-1.0 + (t - 8.0)/2.5)}),
\end{equation}

\begin{equation}
	\mathrm{IMF_{weight}^{'}} = \mathrm{weight} \times \mathrm{IMF_{weight}},
\end{equation}

\noindent where $t$ is log age (yr).  This change keeps the integrated light predictions similar to the original \citet{conroy2010} normalization (see Figure \ref{fig:int_light} and below). This new calibration is the default re-normalization for Padova isochrones in FSPS as of v2.5. The calibration introduced in this work is an improvement from previous versions available in FSPS. 

The dusty isochrones from \citet{girardi2010} are comparable to the models presented in this work. There is a difference in the extent in magnitude that two models span seen at $4.5\mu m$. At $8.0\mu m$, however, the overall shape of the luminosity functions are similar. We compared all the dust schemes made available by the Padova group and found none that made an improvement the magnitude range in the \citet{girardi2010} models. The model LMC luminosity function presented in this work is well-matched to the data with the exception of the bump seen at $7<[8.0]<9$. 

However, while improved from previous versions, the model SMC luminosity function still significantly over-predicts the number of AGB stars. The source of this discrepancy is currently unclear but it could be a problem with the treatment of the AGB evolution in the lower metallicity isochrones. Agreement between the models and data for the SMC in Figure~\ref{fig:lmcsmc} indicates that we are likely not over-predicting the amount of AGB dust per $L_{\rm bol}$ but, again, the CMDs hides information regarding AGB lifetimes which are critical for the luminosity functions in Figure~\ref{fig:lf}.  Given that a similar over-prediction is seen in the \citet{girardi2010} models indicates that this is an issue more fundamental than the dust models.

As a further check, we compare predicted AGB number counts for several galaxies from the ACS Nearby Galaxy Survey Treasury \citep[ANGST][]{dalcanton2009} using the HST WFC3 F160W filter. \cite{melbourne2012} compared the observed number of AGB stars for ANGST galaxies to the predictions by the \cite[][MG08]{marigo2008} isochrones and the \cite[][G10]{girardi2010} modifications to them. We present those, along with our models that include the updated normalization of the AGB lifetimes in Table~\ref{table:agb_counts}. As expected, our model predictions are lower than those of MG08. However, there are still cases where the discrepancy with observations is significant, as noted in \cite{melbourne2012} when adopting the more recent G10 isochrones. Our mean model-data number ratio for AGB stars is 1.75, similar to the value from \citet{melbourne2012} when using the G10 isochrones.

\begin{deluxetable}{llllll}
\tabletypesize{\footnotesize}
\tablecolumns{3}
\tablewidth{0pt}
\tablecaption{Comparison of AGB counts for ANGST galaxies \label{table:agb_counts}}
\tablehead{
\colhead{Galaxy} & \colhead{Observed}  & \colhead{MG08} &  \colhead{G10} &  \colhead{This Work} &  \colhead{$\frac{\mathrm{This Work}}{\mathrm{Data}}$}}\\

DD078 &	273	& 858	& 721	& 724 & 2.65\\
DD082 &	1046	& 2556	& 1122	& 1709  & 	1.63\\
IC2574-SGS &	1504	& 2587	& 1554	& 1867 & 1.24\\
NGC4163 &	640	& 1425	& 677 	& 945 & 1.47 \\
UGC4305-1 & 740 & 1438  &  1113 & 1136 & 1.54\\
UGC4305-2 & 721 & 1532 &  1063 & 1428 & 1.98\\
\end{deluxetable}

\begin{figure*}[t]
	\includegraphics[width=1\textwidth]{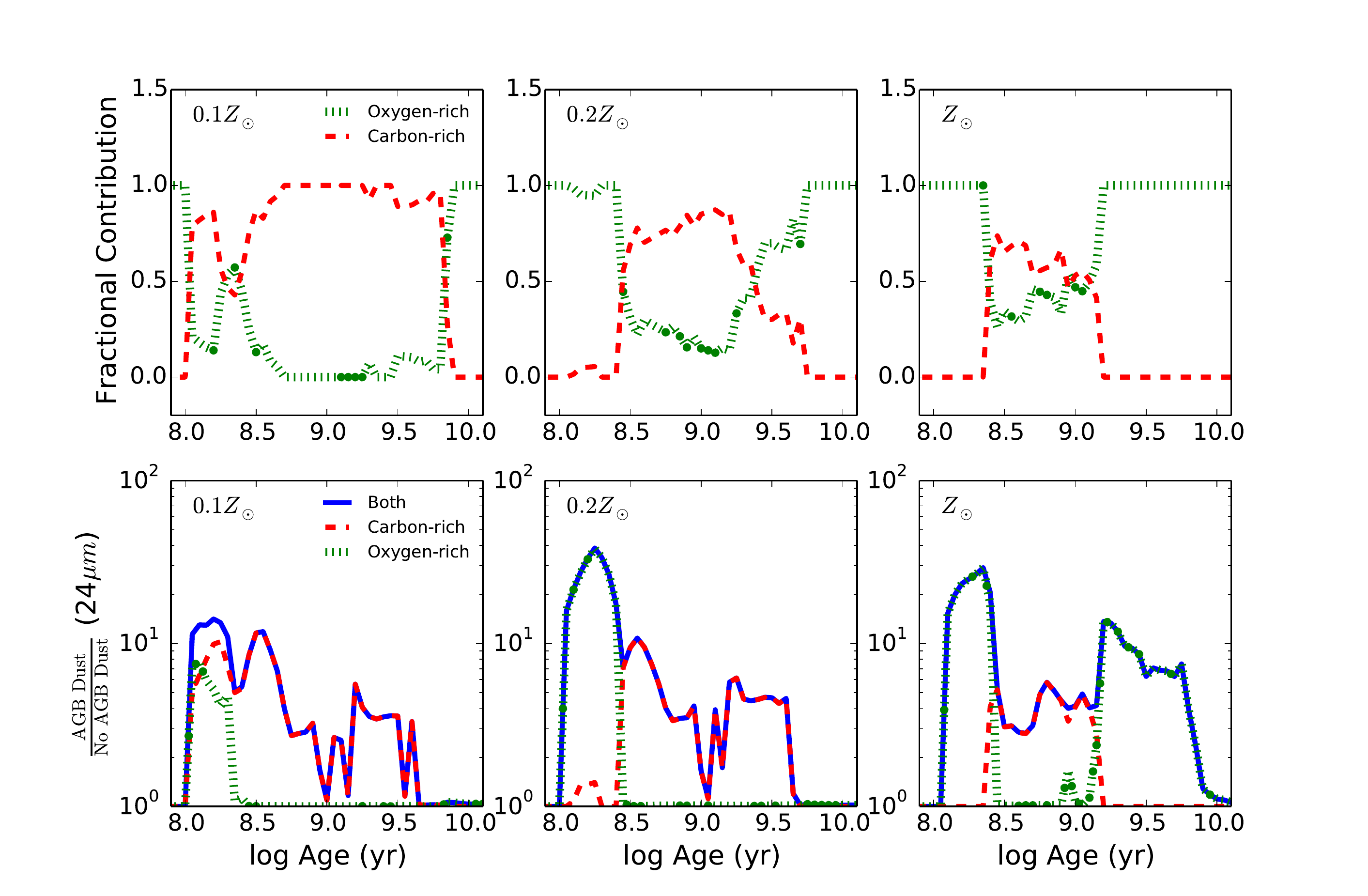}
	\caption{Relative contributions of carbon and oxygen-rich stars to the integrated light of single-age stellar populations. Top Panel: Fractional contribution of carbon and oxygen-rich stars to the total bolometric AGB luminosity.  Bottom Panel: Relative contribution to total $24\mu m$ flux, comparing models with and without AGB dust. This figure highlights both the age and metallicity dependence of carbon-rich star formation and also the disproportional influence of carbon-rich stars to the mid-IR flux when AGB dust is included in the models.}
	\label{fig:star_contrib}
\end{figure*}

Our new normalization of the Padova isochrones was chosen to provide a decent fit to both the LMC luminosity function and the integrated colors of LMC star clusters. The latter is shown in Figure~\ref{fig:int_light}. In this figure we compare both our new normalization and the normalization from CG10 to observations \citep[See Figure 2,][]{conroy2010} . The latter are obtained by stacking star clusters in bins of age to reduce the effects of sparse sampling of the AGB in individual clusters. By construction, the integrated colors are well-matched to the data.

While we have calibrated our models based on Figure~\ref{fig:int_light}, we note that \citet{girardi2013} cautioned about biases introduced due to a ``boosting'' of the number of AGB stars right at the age of many LMC clusters. However, the good agreement between the observed and model LMC LF indicates that the effect of AGB boosting over a narrow range of ages is unlikely to have a significant effect on our results.


We wish to emphasize that our renormalization of the Padova isochrones is completely ad hoc and is simply chosen to provide better fits to observations.  Definitive conclusions regarding AGB lifetimes, etc., will require actual modifications to the underlying physics in the stellar models, as for example done in \cite{girardi2010}.

\section{The Effect of AGB Dust on Integrated Light}

\subsection{General Trends}

In this section we first describe the general conditions in which circumstellar dust may affect the integrated light of simple and complex stellar populations (\S4.1), then move to discussing two examples in which there is evidence for the impact of circumstellar dust (early-type galaxies in \S4.2 and low-metallicity dwarf galaxies in \S4.3), and one example in which circumstellar dust appears to play no important role (actively star-forming galaxies, in \S4.4).

With the new circumstellar dust models we can calculate revised SEDs and compare them to models without circumstellar dust. In Figure~\ref{fig:sed} we show model SEDs with (solid lines) and without (dashed lines) dust shells around AGB stars for ages of 0.2, 2, and 10 Gyr for $Z=0.2Z_{\odot}$ and $Z=Z_{\odot}$.  The circumstellar dust substantially increases the model flux red-ward of $\sim4\mu m$. At old ages the influence of the AGB dust is sharply diminished, having little discernible impact on the SED.  We note here that the models in this figure contain no diffuse dust associated with the interstellar medium (see below).

The trends with age and metallicity in Figure~\ref{fig:sed} can at least partially be understood by considering separately the effects of carbon-rich and oxygen-rich stars on the integrated SED.  We explore this in Figure \ref{fig:star_contrib}, which shows the fractional contribution of oxygen-rich and carbon-rich stars to $L_{\rm bol}$ (top panels) and the relative contribution of oxygen-rich and carbon-rich stars to the $24\mu m$ flux (bottom panels).  Note that we are using Padova stellar models here.

We see the well-known fact that the appearance of carbon-rich stars is both metallicity and age dependent. In the top row of Figure \ref{fig:star_contrib} we see that for the lowest metallicity model, $Z=0.1Z_{\odot}$,  carbon-rich stars dominate $L_{\rm bol}$ for a longer period of time than the higher metallicity models. Moreover, carbon-rich stars do not exist at old ages, $9.3\lesssim {\rm log}\,t/yr \lesssim10$ for high metallicities. The age dependence is complicated, depending on processes such as the TDU and HBB. Detailing when and under what conditions carbon-rich stars are produced is outside the scope of this work but it is a much discussed subject in the literature \citep[e.g.,][and the references therein]{iben1983, dicriscienzo2013}. It is important to understand because at certain ages and metallicities carbon-rich stars are dustier and therefore redder than oxygen-rich stars, as demonstrated in the lower panels of Figure~\ref{fig:star_contrib}.  We see in Figure~\ref{fig:star_contrib} that carbon-rich stars have a disproportionate influence on the $24\mu m$ flux relative to their overall $L_{\rm bol}$ contribution. The lack of carbon-rich stars at old ages, combined with the generally lower AGB luminosities at late times (see Figure~\ref{fig:lum_func}) explains why circumstellar dust emission from AGB stars has a relatively modest effect on the integrated light SED at old ages.

\begin{figure*}
	\includegraphics[width=1.0\textwidth]{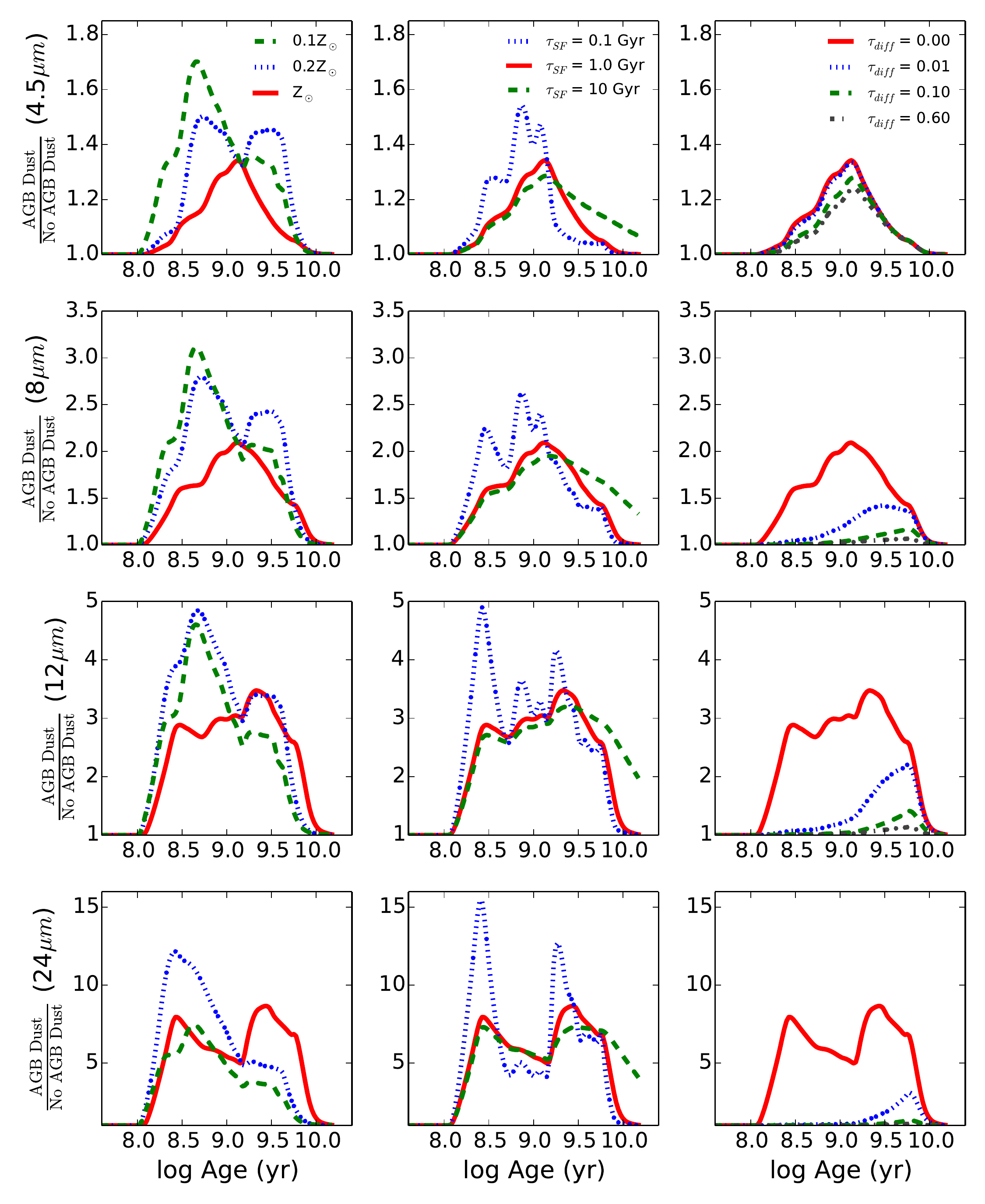}
	\caption{Effect of circumstellar dust on model fluxes as a function of metallicity (left column), star-formation history (middle column), diffuse dust content (right column), and wavelength (top to bottom). The model plotted in red is the default model, same across every panel, with $Z=Z_{\odot}$, $\tau_{\rm SF}$ = 1 Gyr, and $\tau_{\rm diff}=0.0$. The diffuse dust content significantly reduces the influence of AGB circumstellar dust at $\gtrsim8\mu m$ but at $4.5\mu m$ the models are largely insensitive to the presence of diffuse dust. }
	\label{fig:fluxcomp}
\end{figure*}

\begin{figure}[t]
	\includegraphics[width=0.5\textwidth]{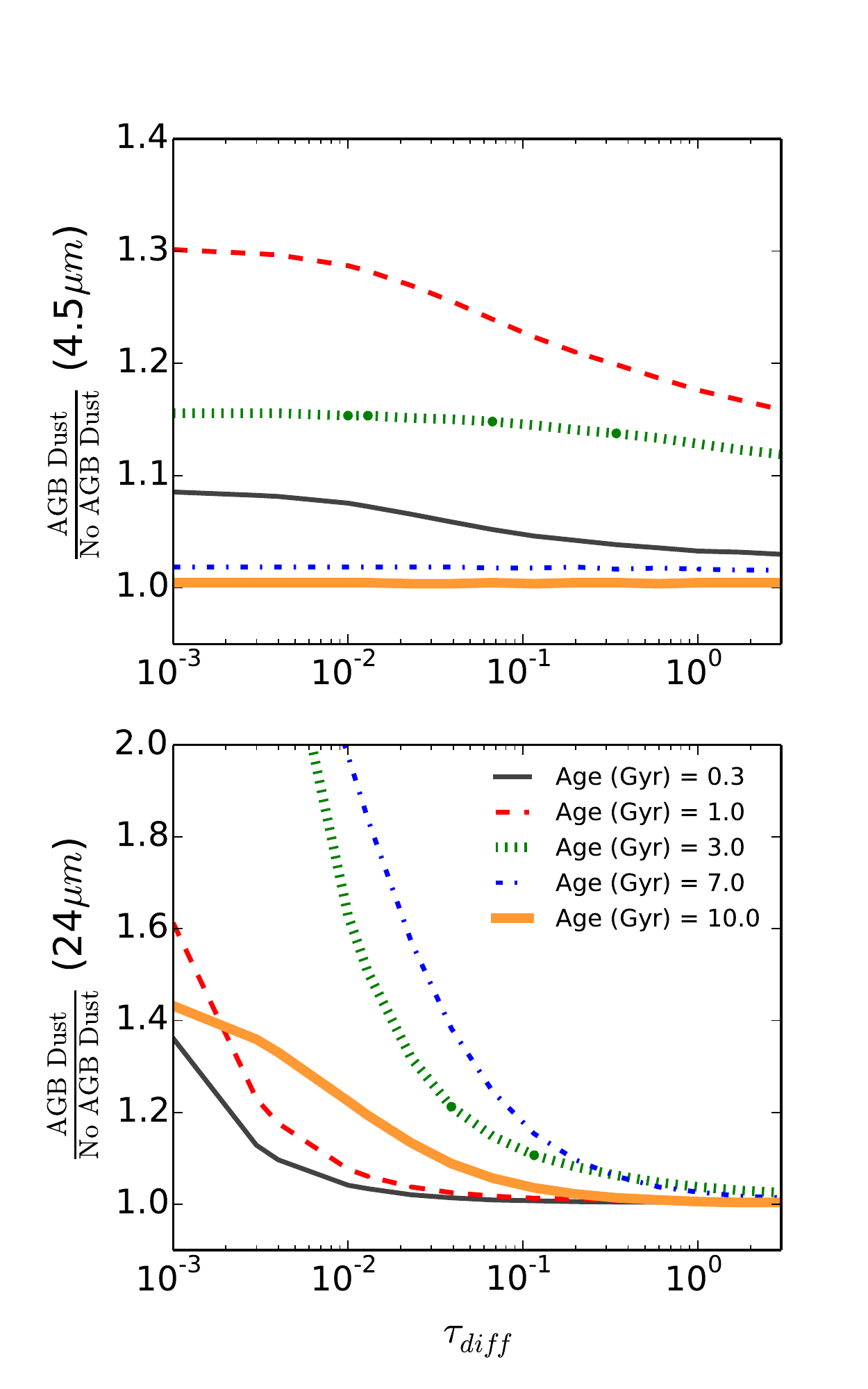}
	\caption{A more detailed look at the effect of diffuse dust on the contribution of AGB dust at $4.5\mu m$ (top) and $24\mu m$ (bottom) for several ages. At $24\mu m$ there is dramatic decline in the influence of AGB dust for even moderate values of diffuse dust. However, flux at $4.5\mu m$ is largely insensitive to the contributions of diffuse dust as the diffuse dust in our model is not hot and therefore does not radiate substantially at such short wavelengths.}
	\label{fig:dust}
\end{figure}

In Figure \ref{fig:fluxcomp} we show how AGB dust affects the NIR and mid-IR integrated light of composite stellar population models as a function of age for different metallicities (note that this refers to populations as a whole, rather than the metallicity of individual stars discussed in \S2.1, which will have an effect on the SPS models), star formation histories (SFHs; represented by exponential models with an e-folding time given by $\tau_{\rm SF}$), and diffuse dust content (the contribution of dust around young stars and dust in the ISM).   In this figure four wavelength ranges are shown and in each panel we compute the ratio of luminosities between models with and without circumstellar AGB dust.

The diffuse dust model is a combination of the \citet{charlot2000} model for dust attenuation, with a power-law attenuation curve (with $\tau_{\rm diff}$ quoted at 5500\AA and the power-law index, $\alpha = -0.7$ unless stated otherwise) and the \citet{draine2007} model for diffuse dust emission. The overall amount of diffuse dust emission is determined by energy conservation. In this paper, models that include diffuse dust also include dust around young stars, the so-called ``birth cloud component'', where $\tau_{\rm bc} = 3 \tau_{\rm diff}$. Note that the \citet{draine2007} dust models include both thermal emission from cold dust and PAH emission. The models are described by three parameters $q_{\mathrm{PAH}}$, $U_{\mathrm{min}}$, $\gamma$. Unless otherwise stated we adopt $3.5\%$, $1.0$, $0.01$ respectively for these values.

We see in all models that only after $\sim0.1$ Gyr does AGB dust become influential, peaking at about 1 Gyr, with a significant reduction by the time the galaxy reaches 10 Gyr in age.  The influence of AGB dust on the SED varies in interesting ways as a function of wavelength, metallicity, and SFH, but by far the most influential parameter is the amount of diffuse dust.

For galaxies where there is no diffuse dust the impact of the dust around AGB stars is significant and extends to wavelengths as short as $\sim4\mu m$. However, even a modest amount of diffuse dust significantly diminishes the relative influence of AGB dust at $\gtrsim8\mu m$. At shorter wavelengths the effect of AGB dust dominates over that of diffuse dust, since in our model the diffuse dust component is generally not hot enough to emit significantly at $<8\mu m$. We investigate this in more detail in Figure~\ref{fig:dust} where we show the effect of AGB dust at 4.5$\mu m$ and 24$\mu m$ as a function of the amount of diffuse dust for several ages. At 4.5$\mu m$ the AGB flux is relatively insensitive to the diffuse dust content but at 24$\mu m$ the diffuse dust greatly reduces the influence of the emission from the AGB dust shells. From $\sim1.0-10.0$ Gyr the AGB dust has a significant influence of the mid-IR light only for low diffuse dust values and, consistent with the previous figures, drops off as the diffuse dust becomes more prominent.  We therefore conclude that circumstellar dust around AGB stars will be important in regimes where the diffuse dust optical depth is roughly $\lesssim0.1$.

\begin{figure}[t]
	\includegraphics[height=0.80\textheight]{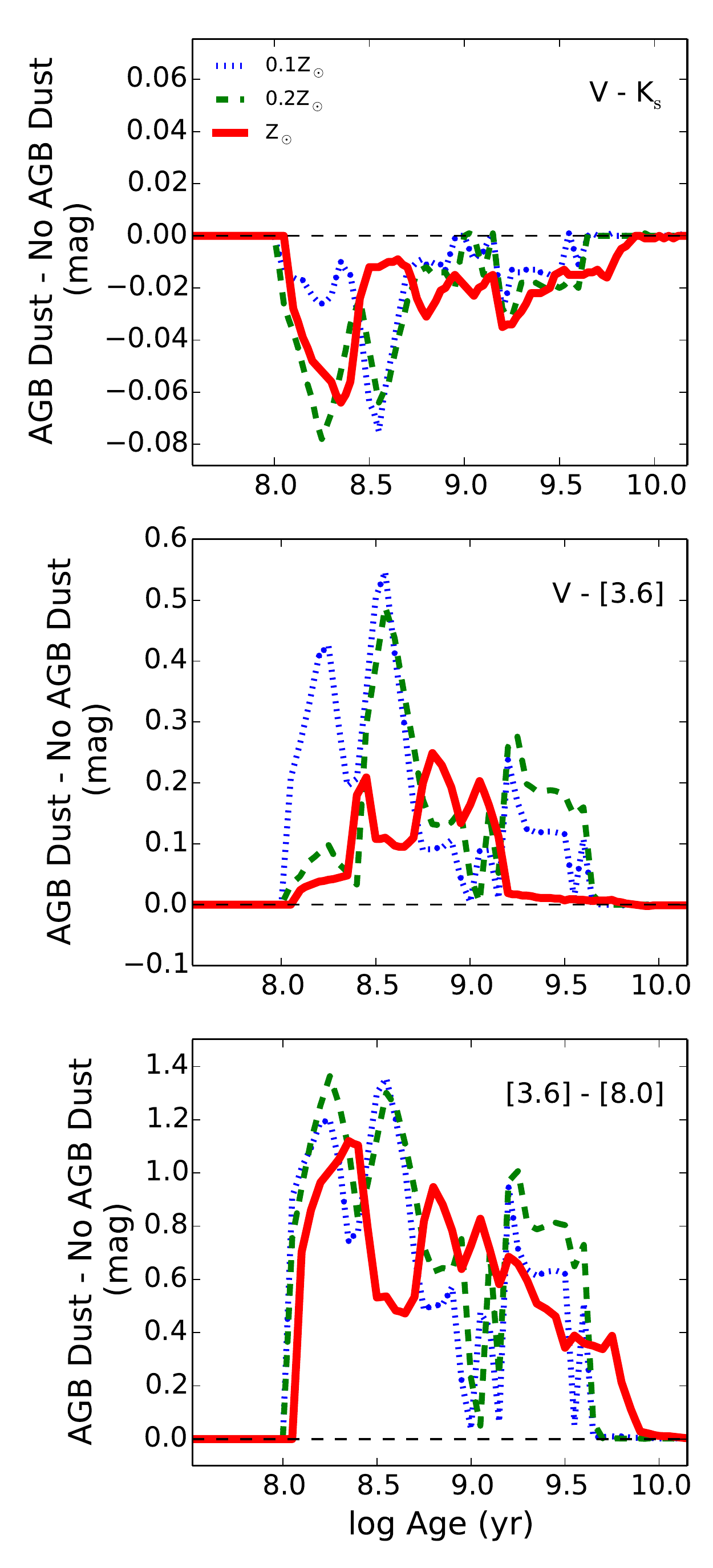}
	\caption{Effect of AGB dust on optical and NIR colors. The effect on mid-IR colors is large and very time- and metallicity-dependent.  The contribution of circumstellar AGB dust to V$-$K$_\mathrm{s}$ is small compared to the other colors, but V$-$K$_\mathrm{s}$ is frequently used in high-precision SED modeling of galaxies and, with the typically small measurement uncertainties of optical-NIR colors, the relatively small effect of circumstellar dust on V$-$K$_\mathrm{s}$ may in fact have a significant effect on parameters derived from SED modeling.}
	\label{fig:colors}
\end{figure}

We also investigate the effect of AGB dust on optical-NIR and NIR-mid-IR colors in Figure \ref{fig:colors}.  The difference is not large for $V-K_{\rm s}$, but this color is commonly employed in high-precision SED modeling of galaxies and so even a small change could potentially induce large differences in derived parameters. The models that include AGB dust tend to be bluer in $V-K_{\rm s}$ than models that do not for a range of metallicities and ages. This is because AGB stars, with or without dust shells, contribute essentially zero flux in the $V$ band and significant flux in $K_{\rm s}$.  Dusty AGB stars can be optically thick even in the $K_{\rm s}$ band, resulting in a lower flux and hence bluer $V-K_{\rm s}$ color.   In the middle and bottom panels of Figure \ref{fig:colors} we see that the models with AGB dust are significantly redder than models without for optical-mid-IR and mid-IR-mid-IR colors.

\subsection{Comparison To Early-Type Galaxies}

Early-type galaxies are potential candidate systems where the effect of AGB dust may be important, owing to their generally low diffuse dust content.  They are known to host intermediate and old age populations, $1<t<10$ Gyr, based on optical absorption line spectroscopy \citep[e.g.][]{trager1998, thomas2005, jimenez2007, trager2005, thomas2005, conroy2014}. While many early-type galaxies do show evidence for diffuse dust at low levels \citep[e.g][]{lauer2005}, the origins of which are actively debated, there exists galaxies with only upper limits in the FIR, suggesting that they have very little diffuse dust \citep[e.g.][]{jura1987, knapp1989, temi2009, martini2013}.

\begin{figure}[t]
	\includegraphics[width=0.5\textwidth]{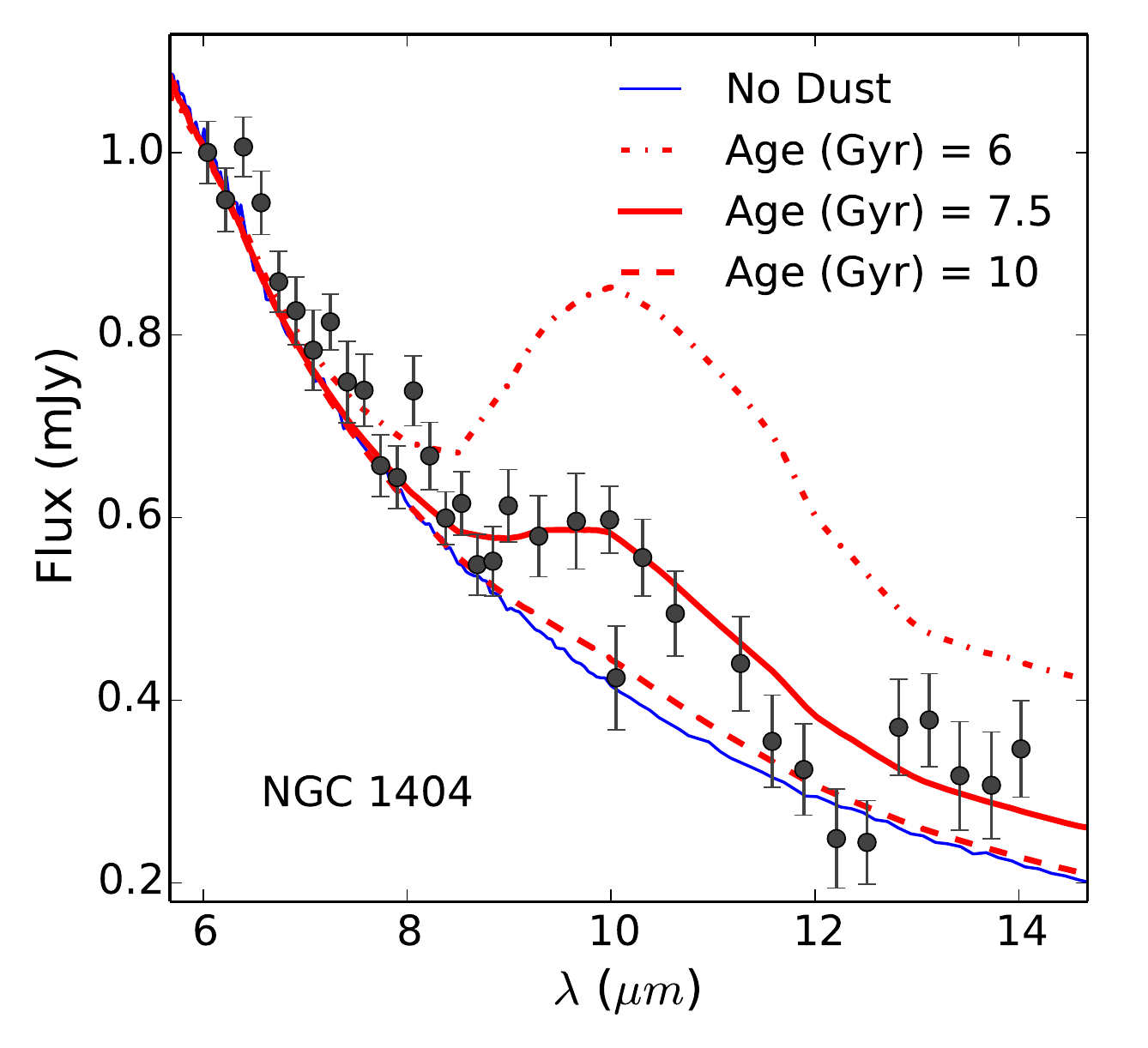}
	\caption{ISO spectroscopy for the early-type galaxy NGC 1404 (points with errors), compared to models with (red) and without (blue) AGB dust. The model with AGB dust gives a much better fit to the data at $>8\mu m$. Moreover, it is clear that the mid-IR is very sensitive to age even over the interval $6-10$ Gyr.}
	\label{fig:athey}
\end{figure}

\cite{athey2002} presented a mid-IR spectrum from the Infrared Space Observatory (ISO) for the early-type galaxy NGC 1404. These authors found a clear excess in the data at $8-14\mu m$ compared to a blackbody extrapolation. They attributed this excess to O-rich AGB features. In Figure~\ref{fig:athey} we show their data for NGC 1404 compared to our models with no dust (blue line) and three models with AGB dust for different ages (red lines). All models assume the galaxy has a metallicity of $Z_{\odot}$.

First, we see that by including AGB dust in the model the fits to the data are greatly improved at $8-12\mu m$. Furthermore, we see from the three models with AGB dust the mid-IR can be a powerful age discriminator, as already discussed in \cite{bregman2006}, \cite{bressan1998}, and \cite{athey2002}, for simple systems with only one burst of star-formation. We find that a model with AGB dust and an age of $\sim$7.5 Gyr provides the best fit to the $10\mu m$ feature. This is in good agreement with the 9 Gyr estimated by \cite{bregman2006}, who used the AGB dust models from \citet{piovan2003} to estimate galaxy ages based on mid-IR SEDs. This age is also in agreement with an optical spectroscopically-estimated age for this galaxy of 7 Gyr (Conroy \& van Dokkum, in prep). 

Interpretation of the mid-IR data for NGC 1404 is complicated by the fact that this galaxy is apparently not entirely devoid of diffuse dust. It is detected by Herschel at $70\mu m$ and $160\mu m$ \citep{temi2009}, indicating the presence of cold dust.  However, we attempted to fit the data from \citet{athey2002} with only diffuse dust (no AGB dust) and found that such a model could not simultaneously fit the mid-IR and FIR data (see also Figure \ref{fig:fir_martini}). Therefore, we conclude that circumstellar AGB dust appears to be necessary to fit the mid-IR data for this galaxy.

\begin{figure*}[t!]
\center
	\includegraphics[width=0.9\textwidth]{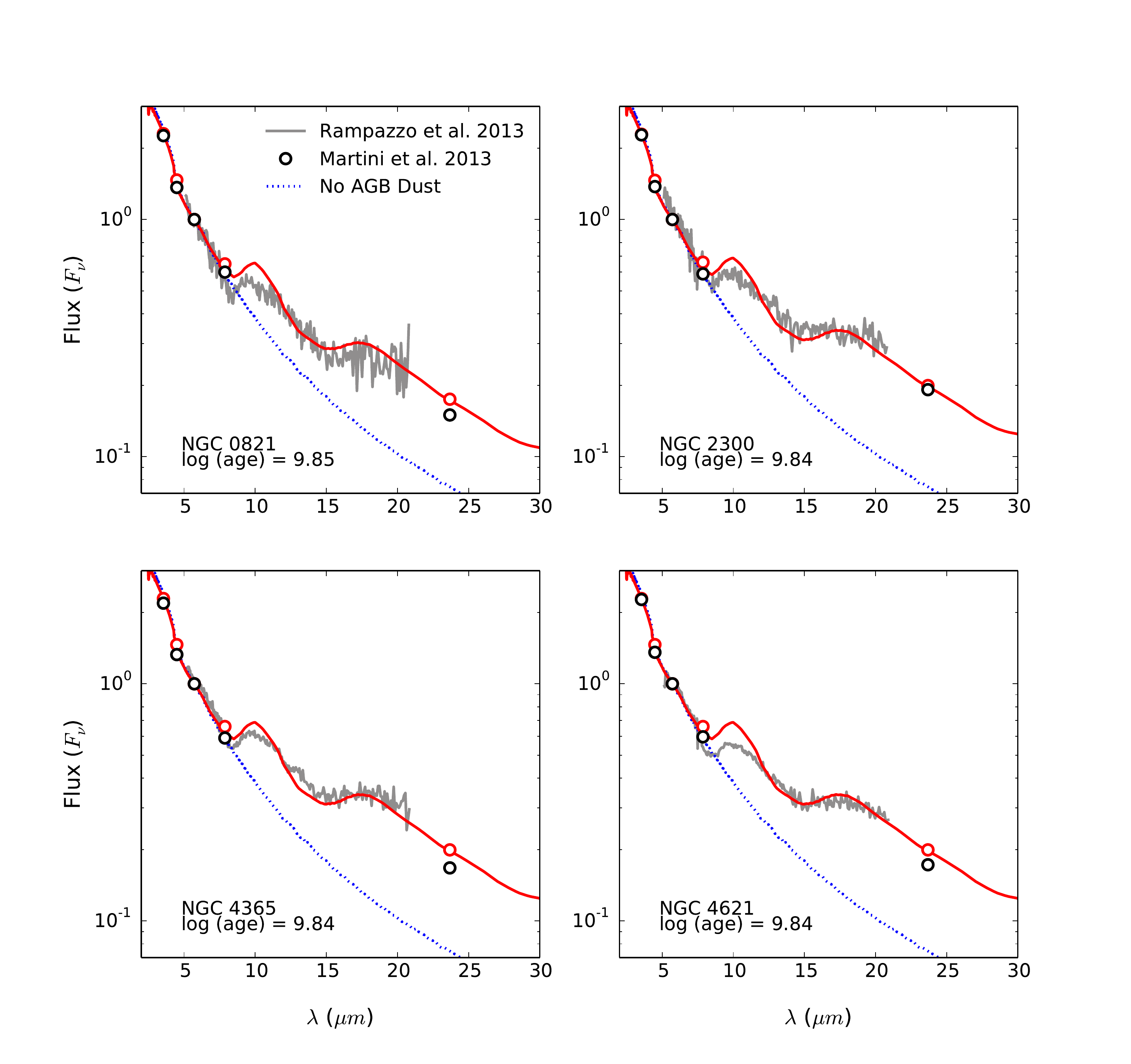}
	\caption{{\it Spitzer} IRAC and MIPS photometry and IRS spectroscopy of a sample of early-type galaxies compared to Z=Z$\odot$ models with (red) and without (blue-dashed) AGB dust. Photometry are from \citet{martini2013} and spectra are from \citet{rampazzo2013}. All data were normalized to 5.72 $\mu m$, the IRAC 3 central wavelength. The errors in the photometry are smaller than the points. The ages were determined through by-eye fits. The models that include AGB dust generally fit the data well, with slight over predictions of the the 10$\mu m$ emission feature seen in all the spectra.}
	\label{fig:martini}
\end{figure*}

\cite{martini2013} recently published mid-IR photometry of early-type galaxies from {\it Spitzer}'s IRAC and MIPS bands. \cite{martini2013} divided the sample based on whether there were diffuse dust lanes detected in HST imaging and detected at 70 and 160 $\mu m$. \citet{rampazzo2013} also published an atlas of mid-IR spectroscopy from  {\it Spitzer}-IRS of early-type galaxies that they sort into classifications based on the contribution of old stellar populations to the mid-IR continuum, where class 0 are galaxies where the old stellar populations are most significant.

\begin{figure}[t!]
\center
	\includegraphics[width=0.5\textwidth]{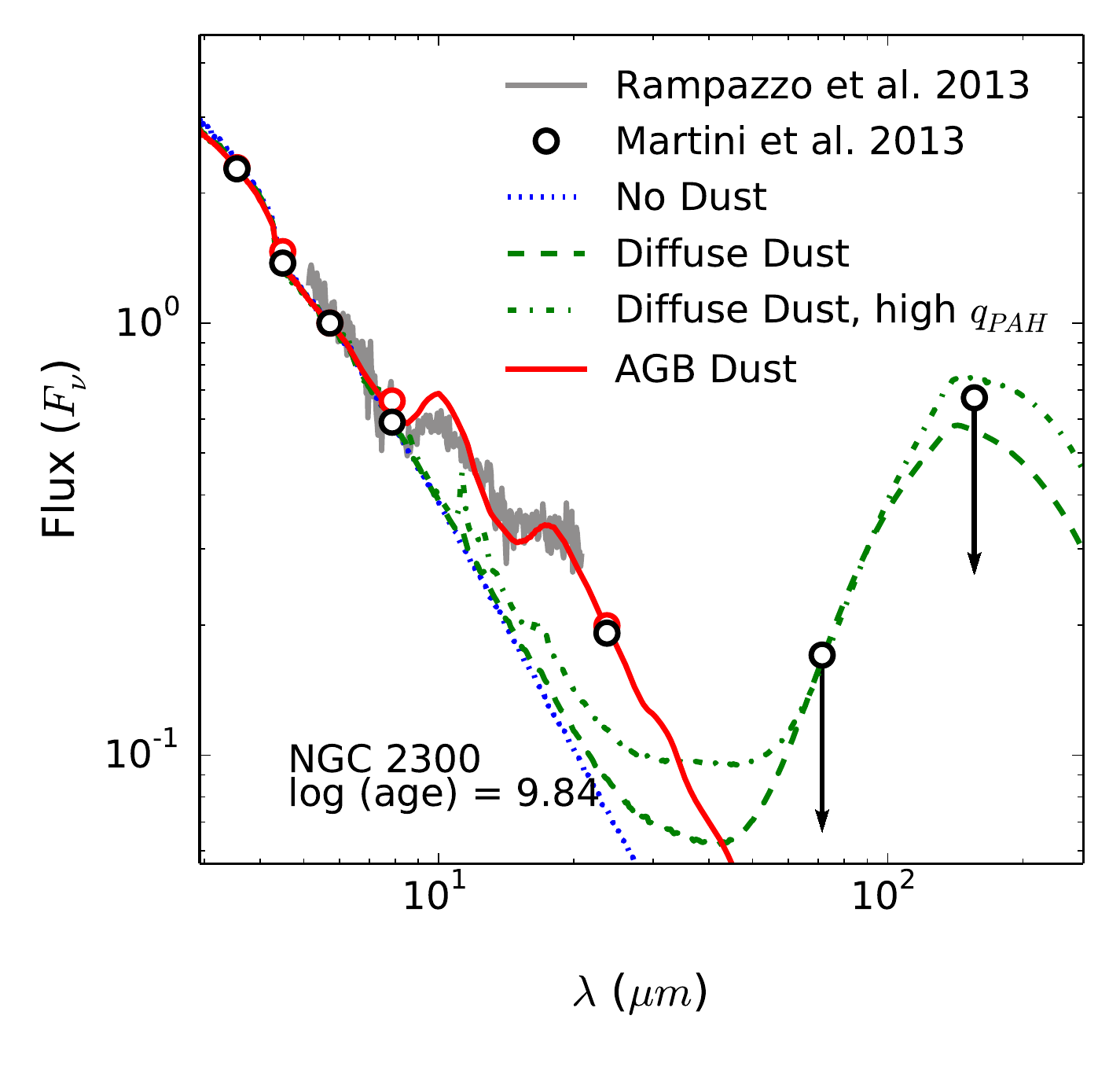}
	\caption{Same as the upper-right panel in Figure \ref{fig:martini} but with two models with only  $\tau_{\rm diff}$ included. The dashed represents the the standard diffuse dust model with $\tau_{\rm diff} = 0.001$. The dot-dashed line represents an otherwise standard model but with the $q_{\rm PAH}$ parameter set at the highest allowable value, 10\%, and  $\tau_{\rm diff} = 0.003$.  The diffuse dust values were chosen to be as large as possible without exceeding the constraints placed $70\mu m$ and $160\mu m$ upper limits. It shows that diffuse dust does not fit the 6-12$\mu m$ features seen in galaxies like NGC 2300 i.e. early-type galaxies with dominant old age stellar populations.}
	\label{fig:fir_martini}
\end{figure}

In Figure \ref{fig:martini} we compare the data for galaxies that did not have detected dust lanes and were classified as class 0 in \cite{rampazzo2013} to models that include (solid-red) and do not include (dashed-blue) AGB dust. All models plotted were made for a solar metallicity, single-age stellar population. The data were normalized to the flux at $5.72 \mu m$, which should eliminate aperture size issues between IRAC and IRS. The ages of the galaxies were estimated by hand to provide a satisfactory fit to the $3-20\mu m$ data. The model that includes AGB dust fits the $3.6-8.0\mu m$ data just as well as the model that does not, only for $\gtrsim 8.0\mu m$ are there significant differences between models with and without circumstellar dust (for these ages). The models provide a good fit to the mid-IR spectra from {\it Spitzer}-IRS. In particular, the overall shape, from $5-20\mu m$, is reproduced well and the relative strength and positions of the broad 10 and $18\mu m$ dust features are generally well matched. 

In Figure \ref{fig:fir_martini} we compare the observations of NGC 2300 with both the AGB dust model and two models that includes only diffuse dust. The fundamental difference between the latter two models is the adopted value of $q_{\rm PAH}$. In one case we adopt the fiducial value while in the other we adopt an unreasonably high value of 10\% \footnote{This is an extrapolation of the \citet{draine2007} models, the highest value \citet{draine2007} compute is 4.58\%. }. We added as much diffuse dust into the models as possible without exceeding the constraints of the FIR flux at $70\mu m$ and $160 \mu m$. It is clear that the model with only diffuse dust cannot reproduce the $8-24\mu m$ features seen in early-type galaxies with influential old stellar populations. Varying the diffuse dust parameters $q_{\mathrm PAH}$ and $U_{\mathrm min}$ does not qualitatively improve the fits.

The slight over prediction of the strength of the $10\mu m$ emission feature is likely due to our treatment in the dust grain properties. We reiterate that our grain scheme is relatively simplistic and there is an abundance of evidence that the actual grain mixtures in AGB stars, and their dependence on the properties of the star, are more complicated than represented in the models presented here \citep[e.g.][]{jones2012, mcdonald2010}. Other groups producing similar dust shell models have incorporated more detailed approaches to the grain schemes. For example, \cite{cassara2013} detail a sophisticated approach to the grain mixtures that accounts for changes in $\tau_{AGB}$ and $\dot{M}$ and \citet{groenewegen2006} include aluminum oxide dust along with the more standard silicate in oxygen-rich AGB stars. In future work we will explore the effect of different grain species on the mid-IR spectra of early-type galaxies.

\subsection{Comparison to Low-Metallicity Galaxies}

\begin{figure}[t!]
	\includegraphics[width=0.5\textwidth]{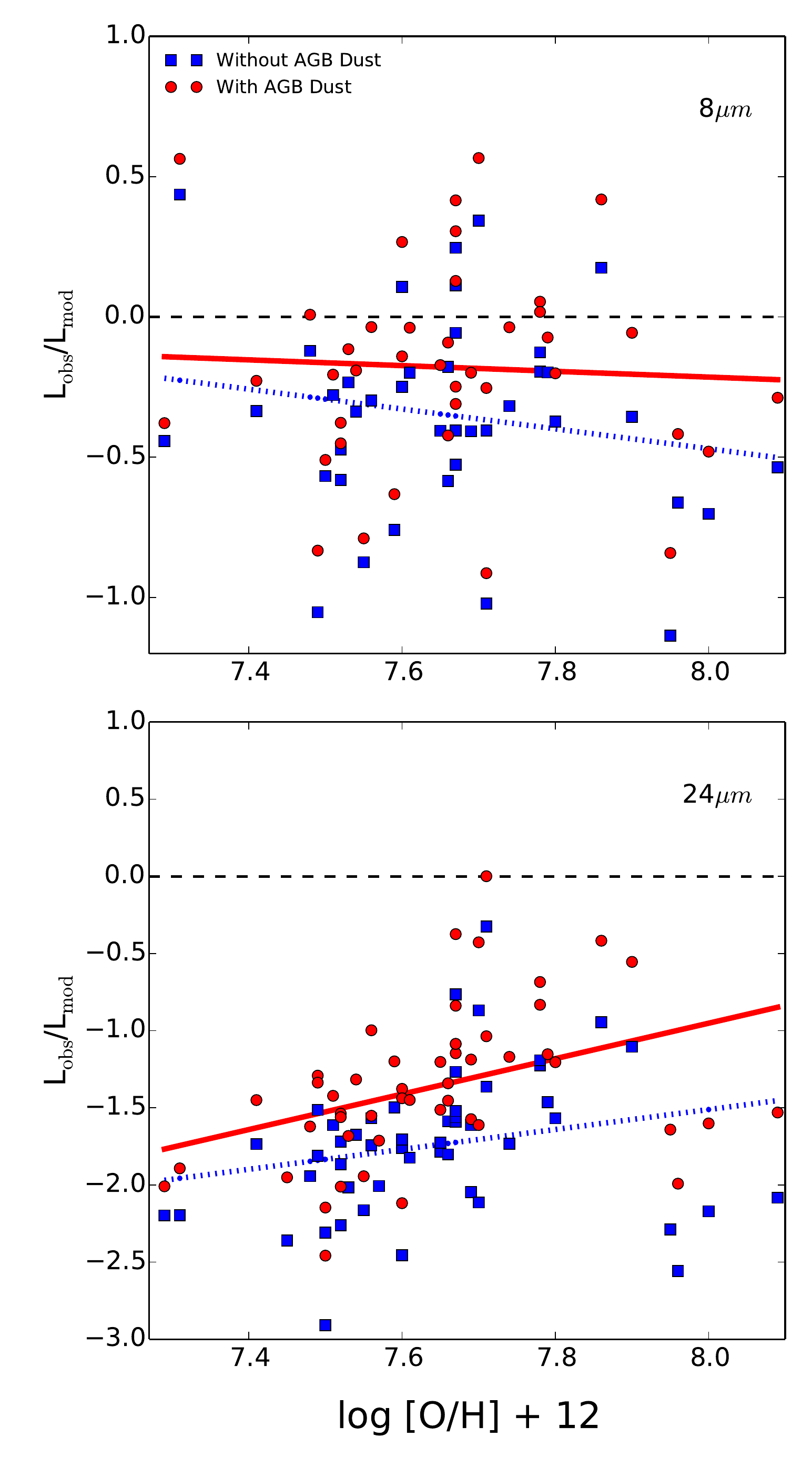}
	\caption{Comparison of the differences between observed and model magnitudes for low-metallicity dwarf galaxies from the ANGST survey.  Results are shown for models both with and without circumstellar dust around AGB stars.  The integrated light model predictions utilize the CMD-based SFHs for the ANGST galaxies.  The lines plotted in each panel are linear fits to the points. The average difference between observed and model magnitudes is nearly zero for the $8\mu m$ flux but only slightly reduces the tension at 24$\mu m$.}
	\label{fig:angst}
\end{figure}

Low-metallicity galaxies are another class of objects in which circumstellar dust may significantly contribute to the integrated SED owing to their generally low diffuse dust content. \cite{johnson2013} compared observations for a subset of ANGST galaxies with SPS models. They found and discussed the possible causes of a 0.2 dex under-prediction by the models compared to  {\it Spitzer} data from the LVL survey presented in \citet{dale2009} \citep[see Figure 5 in][]{johnson2013}. They considered the possibility that circumstellar dust around AGB stars, not included in the SPS models, may be the source of the discrepancy.

\begin{figure}[h!]
	\includegraphics[width=0.5\textwidth]{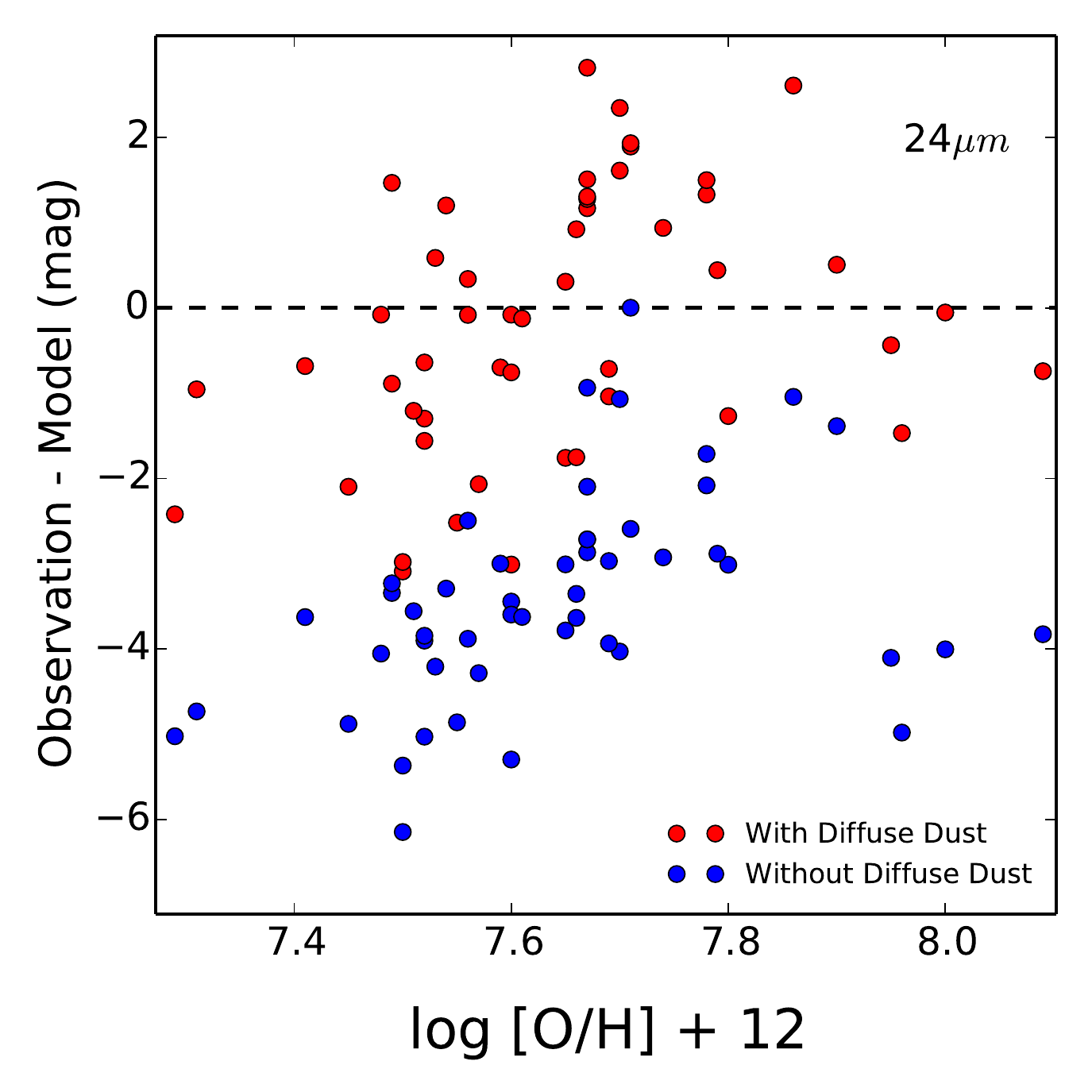}
	\caption{Comparison of the residuals between the predicted and observed $24\mu m$ magnitude for the low-metallicity ANGST galaxies as a function of metallicity. The addition of diffuse dust  ($\tau_\mathrm{diff} = 0.05$, red points) helps resolve the residuals that persist after including only circumstellar AGB dust (blue points), indicating that diffuse dust is important in these galaxies.}
	\label{fig:angst_residual}
\end{figure}

\begin{figure*}[t!]
	\includegraphics[width=1.0\textwidth]{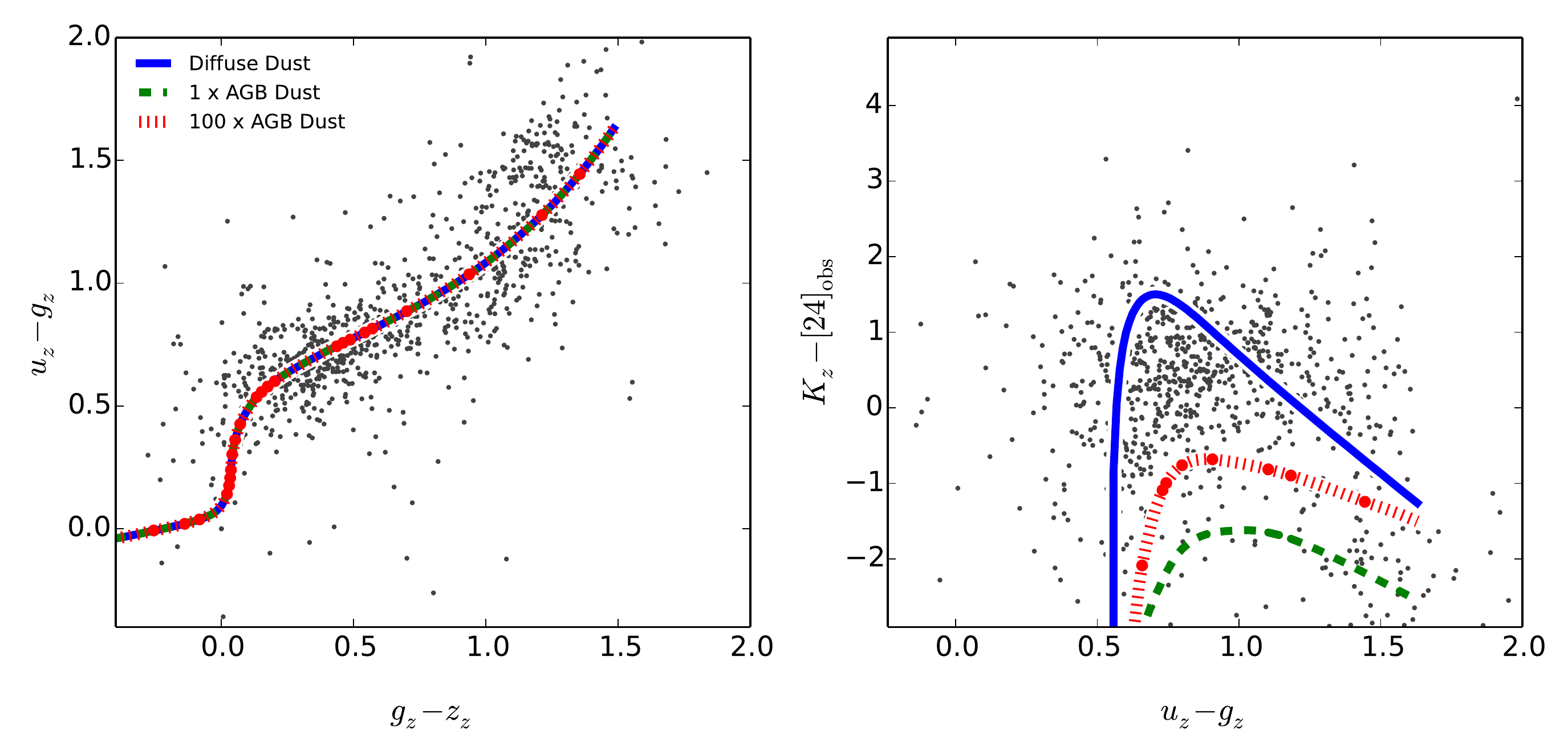}
	\caption{Color-color plots of galaxies at $z\sim1$ from \cite{wuyts2008}. The data are compared to stellar population models with $\tau_{\rm SF}=2.0$ Gyr. Three models are plotted in each panel: a model with only diffuse dust (solid blue; $\tau_{\rm diff}=0.05$), a model with the default amount of AGB dust (dashed green), and a model with the AGB dust optical depth increased by a factor of 100 (red dotted).  Along each line the model age varies from 10 Myr to 13 Gyr.  The models without diffuse dust are unable to match the locus of mid-IR colors for the bulk of the galaxies \citep[compare with][]{kelson2010}.}
	\label{fig:kelson}
\end{figure*}

In Figure~\ref{fig:angst} we plot the offset between observed IRAC 8$\mu m$ and MIPS $24\mu m$ magnitudes and predicted magnitudes for models with (blue squares) and without (red circles) AGB dust as a function of metallicity. For about half the galaxies, the metallicities are well constrained by nebular emission lines \citep[][]{berg2012}. For the rest of the sample, metallicities are estimated the mass-metallicity relationship \citep[][]{berg2012, lee2006}; the resulting metallicities that are broadly consistent with those estimated from HST CMDs \citep[][]{weisz2011}. For each galaxy, the corresponding model has the same SFH and metallicity. The model photometry for each galaxy is based on the estimated SFH for each galaxy derived from CMDs; see \citet{johnson2013} for details.  The default models do not contain any diffuse dust.

In both panels we include linear fits to both sets of offsets; in general the offset is reduced with the inclusion of AGB dust. For the $8.0\mu m$ flux, the average difference between observations and models is close to zero. However, for the $24\mu m$ flux, while there is some improvement in the offset in all the galaxies, a significant discrepancy remains even after including AGB dust. 

We see a very strong correlation between metallicity and the size of the offsets at $24\mu m$. The origin of the correlation is unclear to us. In Figure \ref{fig:angst_residual} we compare the residuals between the observations and models that only include circumstellar AGB dust and models that include AGB dust and diffuse dust at $24\mu m$ ($q_{\mathrm{PAH}} = 0.05$ and dust attenuation index $\alpha = -1.3$). We see that by including diffuse dust we are able to better capture the $24\mu m$ flux. However, the correlation with metallicity remains. This indicates that joint modeling of UV-IR SEDs is necessary for these galaxies to fully understand their characteristics.


\subsection{Comparison to Star-Forming Galaxies at z$\sim$1}

Finally, we consider the effect of circumstellar dust on the SEDs of actively star-forming galaxies. 

\cite{kelson2010} suggested that AGB stars may be responsible for a substantial fraction of the mid-IR luminosities in galaxies from $z=0$ to $z=2$. In Figure \ref{fig:kelson} we show color-color diagrams of galaxies at $z\sim1$ from the FIREWORKS catalog, see \citet{wuyts2008} for details \citep[also, see Figure 2 from][]{kelson2010}.  In each panel we include three models: a model with the standard amount of AGB dust (``$1\times$ AGB dust'') where the amount of AGB dust included in the model is as described in \S2.2; a model with a much larger amount AGB dust (``$100\times$ AGB dust'') where we multiplied the standard dust optical depth by 100, and a model with no AGB dust but a modest amount of diffuse dust ($\tau_{\rm diff}=0.05$). Each model assumes $Z=Z_{\odot}$, $\tau_{\rm SF} = 2.0$ Gyr, and uses a delayed $\tau$-model to be as similar as possible with the models from \citet{kelson2010}. In the first panel we see no difference among the different models. This is expected since only optical colors are plotted and no dust will effect the flux at these wavelengths. This plot serves as a check that the models we have adopted are a reasonable description of the data and that they span the full range of optical-optical colors.

The second panel is an optical-mid-IR color-color plot and we see significant differences among the models. The standard AGB model is able to reproduce only a small fraction of the galaxy colors but fails to span the whole extent of the color range. Even when boosting the AGB dust optical depth by a factor of 100 (far beyond what is plausible) we are unable to reproduce the full color range of the data. Furthermore, reducing the metallicity of the model to $Z=0.4Z_{\odot}$ and thereby increasing the number of carbon stars (see Figure~\ref{fig:star_contrib}), and hence the MIR flux, only modestly improves the fits of the model to the data. However, a modest amount of diffuse dust can readily explain the full range of mid-IR colors. It therefore seems unlikely that AGB dust alone can explain the full range of the mid-IR colors.  

This issue has also been discussed in \cite{melbourne2013} where they compared the mid-IR colors of AGB stars in the SMC and LMC with the colors of AGB stars adopted for the models presented in \cite{kelson2010}. \citet{melbourne2013} concluded that the average AGB color adopted by \citet{kelson2010} were significantly redder than the observed average AGB colors. This is due to the fact that \cite{kelson2010} used heavily dust enshrouded AGB stars as representative of more typical AGB stars.  Furthermore, \citet{kelson2010} used the \citet{maraston2005} stellar population models, which appear to assign too much weight to the TP-AGB phase \citep[][]{kriek2010, zibetti2012}.

The results of this section are consistent with our findings from \S4.1 as well as the findings of \cite{melbourne2013}. Actively star-forming galaxies generally contain significant quantities of diffuse dust that tends to overwhelm the effect of circumstellar dust in the mid-IR.

\section{Discussion}

In this work we computed a set of empirically-calibrated radiative transfer models for the dusty circumstellar shells around AGB stars.  These models were coupled to a stellar population synthesis code in order to explore under what circumstances AGB dust might impact the integrated SEDs of composite stellar populations (i.e., star clusters and galaxies).  The models indicate that AGB dust will play an important role in shaping observed SEDs of both quiescent and star-forming galaxies when $4\mu m\lesssim\lambda\lesssim 8\mu m$.  At longer wavelengths, the influence of AGB dust will be substantial only when the diffuse dust content is low ($ A_{\rm V} \lesssim 0.1$).  These conclusions are in good agreement with the results of \citet{silva1998}, who included both circumstellar dust around AGB stars and diffuse dust in their SED models.

Modeling the contribution of dusty circumstellar shells to the integrated light of stellar populations carries a number of uncertainties that are presently difficult to quantify.  These include the lifetimes, luminosities, mass-loss rates, and compositional changes of AGB stars, the structure, extent, and lifetime of the circumstellar shell, the types and quantity of dust within the shell, and the optical properties of dust grains.  We attempted to calibrate and/or mitigate these uncertainties by tuning the model to match several key observables include the SEDs of oxygen and carbon-rich stars, the morphology of the color-magnitude diagram of the LMC and SMC, the luminosity function of stars in the LMC and SMC, and the numbers of AGB stars in low metallicity galaxies from the ANGST survey.  

The luminosity functions proved to be a strong test of the AGB lifetimes in the Padova stellar models.  The default Padova models produce too many stars even at $4.5\mu m$ where dust emission plays a minor role, suggesting that the circumstellar dust models are not the source of the discrepancy. After adjusting the lifetimes downward (by lowering their weight in the synthesis) we found good agreement for both the $4.5\mu m$ and $8.0\mu m$ LFs for the LMC.  However, the model SMC LF still substantially overpredicts the observed counts by factors of $3-5$. Moreover, our recalibration of the Padova models produces relatively good agreement with AGB counts of low metallicity ANGST galaxies, with a typical overprediction of only 1.5 \citep[in agreement with][]{girardi2010}.  This indicates that the Padova isochrones are doing well both at low metallicities probed by the ANGST galaxies ( $Z < 0.15 Z_{\odot}$) and LMC metallicity, but oddly not at SMC metallicity. The \citet{girardi2010} dusty isochrones show a similar overprediction of the observed SMC luminosity function. This suggests that there is a problem with either the SMC IRAC LF, our adopted star formation history for the SMC, or perhaps there is some unknown issue with the models unique to SMC-like metallicities and ages.

There is a well-known degeneracy between stellar age and metallicity when modeling optical data that is typically only broken when the strengths of both balmer and metal lines can be accurately measured.  \citet{bressan1998} argued that circumstellar dust emission around AGB stars might also help to break this degeneracy, since the contribution of AGB stars to the bolometric luminosity varies strongly with time.  We confirm the potential power of mid-IR emission as an age diagnostic by fitting ISO spectroscopy of NGC 1404, a galaxy with no on-going star formation.  The models strongly prefer an age of 7.5 Gyr and are capable of discriminating between even $1-2$ Gyr differences in age at old times.  Remarkably, the age derived in this way agrees very well with the analysis of the optical spectrum for this galaxy.  The key limiting factor to deriving ages based on mid-IR data is systematic uncertainties in the modeling.  Reducing these systematic uncertainties with targeted observations of AGB stars in the Local Group would enable firm absolute ages to be obtained in the mid-IR, even for old stellar populations. 

With the IRAC and MIPS $24\mu m$ photometry and {\it Spitzer}-IRS spectroscopy presented by \citet{martini2013} and \citet{rampazzo2013}, respectively, we constructed a small sample of quiescent, early-type galaxies lacking evidence for cold dust. The models presented here provide an reasonable fit to the data. This indicates that the majority of the mid-IR emission in these galaxies is dominated by circumstellar dust, as also argued by \citet{martini2013}. In detail, the models slightly over predict the $10\mu m$ silicate feature, which may be a reflection of our overly simplistic treatment of the grain species in the circumstellar shells. 

\cite{johnson2013} used CMD-based star formation histories to make predictions for the integrated light of nearby dwarf galaxies.  Comparison of these predictions to observed total fluxes revealed good agreement in the FUV$-$NIR, but an offset was seen at $8\mu m$.  We repeated this test with our updated models and found that the offset is reduced from 0.33 dex to 0.13 dex when circumstellar dust is included.  We also compared the models to the observed $24\mu m$ flux and found that the offset without circumstellar dust, which is greater than a factor of 30, is only slightly reduced when circumstellar dust is included.  It is likely that low levels of diffuse dust in these systems is providing the bulk of the observed $24\mu m$ emission.

There has been much discussion in the literature regarding the sources responsible for the mid-IR emission in typical star-forming galaxies.  Possibilities include dust heated by young stellar populations, dust heated by old stellar populations, and circumstellar dust around AGB stars.  \cite{kelson2010} suggested that the latter could play a large role in the mid-IR emission of star-forming galaxies.  This possibility is not borne out in our models.  Instead, circumstellar dust provides a rather small contribution to the mid-IR flux for galaxies with typical amounts of diffuse dust ($A_{\rm V}\gtrsim0.1$).   This result does not preclude the possibility that evolved stars, including AGB stars, contribute substantially to the heating of the diffuse dust.  Indeed, \cite{salim2009} argued for this based on observations of star-forming galaxies at $z\sim1$ \citep[see also][]{utomo2014}.

The launch of JWST will turn the spotlight on the $\sim1-30\mu m$ SEDs of stellar populations across time.  In this wavelength interval a variety of distinct physical processes contribute to the light, including PAH emission and circumstellar dust.   Models such as the one presented in this paper will be necessary to interpret this data.  In addition, JWST will provide stringent tests of these models by observing nearby AGB stars in detail and entire populations of such stars in nearby galaxies.  In the process we hope to learn not only about the evolutionary history of galaxies but also about the physics shaping the final stages in the lives of AGB stars.

\section{Summary}

We now summarize our main results.

\begin{itemize}

\item In agreement with previous work, our models indicate that the dusty circumstellar shells around AGB stars can have a factor of $\sim10$ effect on the integrated flux of star clusters and galaxies at $\lambda\gtrsim4\mu m$.  The effect is especially large at intermediate ages $\sim0.1-3$ Gyr and when the presence of diffuse dust is minimal. The emission associated with circumstellar dust shells may be important for interpreting the SEDs of quiescent galaxies at high redshift, where the average stellar age is relatively young.
	
\item The presence of diffuse dust in a galaxy sharply reduces the influence of circumstellar dust emission at $\gtrsim8\mu m$.  At $24\mu m$, circumstellar dust contributes only $\sim10$\% of the flux for star-forming galaxies with $A_V\sim0.1$.  At shorter wavelengths  (i.e.,$< 10 \mu m$), where under normal conditions diffuse dust does not contribute substantially, circumstellar dust may still play an important role.

\item In early-type galaxies lacking signatures of cold diffuse dust, the mid-IR data is far in excess of a blackbody extrapolation at shorter wavelengths. Models that include circumstellar dust shells accurately reproduce the mid-IR photometry and spectra of early-type galaxies. The model MIR flux is sensitive to age, even at old ages. This suggests that this wavelength range can be a useful age diagnostic once circumstellar dust models have been accurately calibrated. The $10\mu m$ silicate feature is qualitatively well matched but there are some quantitative differences at the $\sim 10\%$ level.

\end{itemize}

\acknowledgements  We would like to thank Martha Boyer, Martin Groenewegen, Dan Kelson, Ivo Labbe, Sundar Srinivasan, Benjamin Sargent, and Roberto Rampazzo for generously sharing their data, and Paul Martini, Dan Kelson, and Julianne Dalcanton for fruitful conversations on this topic. We also thank the anonymous referee for their thoughtful comments that improved the quality of this manuscript. This project was supported in part by NASA grant NNX13AI46G.


\end{document}